\documentclass[preprint2]{aastex}

\usepackage{longtable}

\shorttitle{UV Spectroscopic indices} 
\shortauthors{Chavez et al.}

\begin{document} 

\title{Synthetic Mid-UV Spectroscopic Indices of Stars}

\author{M. Chavez$^{1,2}$, E. Bertone$^1$, A. Buzzoni$^3$, M. Franchini$^4$, M.~L. Malagnini$^{4,5}$, C. Morossi$^4$,}
\and \author{L. H. Rodriguez-Merino$^1$}
\affil{$^1$INAOE, Luis Enrique Erro 1, 72840, Tonantzintla, Puebla, Mexico}
\email{mchavez@inaoep.mx, ebertone@inaoep.mx, lino@inaoep.mx}
\affil{$^2$Instituto de Astronomia y Meteorologia, Universidad de Guadalajara, Av. Vallarta 2602, Col. Arcos Vallarta, C. P. 44130, Guadalajara, Jalisco, Mexico}
\affil{$^3$INAF, Osservatorio Astronomico di Bologna, Via Ranzani 1, 40127 Bologna, Italy}
\email{buzzoni@bo.astro.it}
\affil{$^4$INAF, Osservatorio Astronomico di Trieste, Via Tiepolo 11, 34131, Trieste, Italy}
\email{franchini@ts.astro.it, morossi@ts.astro.it}
\affil{$^5$Dipartimento di Astronomia, Universit\`a degli Studi di Trieste, Via Tiepolo 11, 34131, Trieste, Italy}
\email{malagnini@ts.astro.it} 

\begin{abstract}
Using the UVBLUE library of synthetic stellar spectra we have computed a
set of mid-UV line and continuum spectroscopic indices. We explore their
behavior in terms of the leading stellar parameters ($T_\mathrm{eff}$,
$\log{g}$). The overall result is that synthetic indices follow the general
trends depicted by those computed from empirical databases. Separately we also
examine the index sensitivity to changes in chemical composition, an analysis
only feasible under a theoretical approach. In this respect, lines indices
Fe~\textsc{i}~3000, BL~3096 and Mg~\textsc{i}~2852 and the continuum index
2828/2921 are the least sensitive features, an important characteristic to be
taken into account for the analyses of integrated spectra of stellar
systems. We also quantify the effects of instrumental resolution on the
indices and find that indices display variations up to 0.1~mag in the
  resolution interval between 6--10~\AA\ of FWHM. We discuss the extent to
which synthetic indices are compatible with indices measured in spectra
collected by the International Ultraviolet Explorer (IUE). Five line and
continuum indices (Fe~\textsc{i}~3000, 2110/2570, 2828/2921, S2850, and
S2850L) display a remarkable good correlation with observations. The rest of
the indices are either underestimated or overestimated, however, two of them,
Mg~Wide and BL~3096, display only marginal discrepancies. For 11 indices we
give the coefficients to convert synthetic indices to the IUE 
system. This work represents the first attempt to synthesize mid-UV indices
from high resolution theoretical spectra and foresees important applications
for the study of the ultraviolet morphology of old stellar aggregates.
\end{abstract}


\section{Introduction}

With the accessibility to the rest-frame mid-ultraviolet (2200--3200~\AA)
spectra of distant evolved systems ($0.5 < z < 2.0$), the analysis of the
ultraviolet morphology of intermediate and late type stars and old systems in
the local universe has regained interest (e.g. \citealp{Spinrad97};
\citealp*{Nolan01}; and, more recently, \citealp{McCarthy04};
\citealp{Cimatti04}). Dating galaxies which have gone through null or
negligible star formation events over the past few Gigayears at these redshifts has fundamental implications not
only from an individual object point of view, but also at cosmological scales.

The mid-ultraviolet offers a new spectral window which may help to break
the so-called age-metallicity degeneracy present in the optical spectra of
evolved stellar systems \citep{Worthey94,Buzzoni95,Jimenez04}. In general, the UV
analyses of stellar populations have relied on empirical stellar databases,
mainly the 
one constructed upon IUE low resolution data \citep{Wu83}. Among the most
important efforts to characterize the UV morphology of stars that presumably
dominate the UV spectrum of old systems was presented by \citet*{Fanelli87} and
\citet{Fanelli90,Fanelli92}. They have defined 25 narrow-band indices in the ultraviolet 
covering two spectral segments: 1230--1930~\AA\ and 1950--3200~\AA,
corresponding to the wavelength limits of IUE cameras. Spectroscopic indices defined in the 
far-UV \citep{Fanelli87} were created to provide alternative tools for the 
analysis of active star-forming galaxies. The second set of indices 
\citep{Fanelli90} was constructed with the aim of analyzing the properties of 
prominent spectral features in old populations. In general, these indices
focus on wavelength regions that are more sensitive to stellar atmospheric
parameters and less affected by the IUE instruments artifacts and by the effects of
features (emission and/or absorption) of interstellar origin. Given the prevailing
contribution to the mid-UV wavelength range of bright main sequence stars \citep{OConnell99,Buzzoni02},
the study of this spectral region is of special importance as a diagnostic tool,
for instance to date high-redshift galaxies \citep{Heap98}.

It has been recognized, however, that empirical databases are generally
composed of stars of the solar neighbourhood and therefore unavoidably carry
on the evolutionary imprints of the Milky Way. In order to cope the paucity of
chemical composition of empirical libraries required for the analysis of old
stellar systems, many of the population synthesis works have
incorporated theoretical libraries of stellar fluxes. Amongst the most popular
libraries is the Kurucz grid of spectral energy distributions (SEDs) which
is public through his web site\footnote{http://kurucz.harvard.edu/}.
It has been found that Kurucz flux library well represents segments of the
mid-UV spectra of stars of intermediate and late type, in particular the
continuum measurements (slopes in the SEDs) and some features such as the
blend at about 2538~\AA\ (LFB00). The limitations of the public Kurucz grid of
fluxes has been a matter of discussion in several investigations.
In this respect \citet*{Dorman03} concluded that for the analysis of the
integrated UV properties of stellar aggregates, low resolution synthetic
fluxes should be preferred over high resolution data. The reason being that use
of high resolution theoretical data is precluded by the still unsolved
uncertainties in atomic and molecular line data. It is important to note,
nevertheless, that Kurucz grid of fluxes have a wavelength sampling of 10~\AA\
and therefore might lack of the appropiate resolution for a direct comparison
with, for example, the commonly used IUE database.

While we agree that, for now, only low resolution (e.g. broad band indices)
can be safely used, it is important to stress that at low resolution a great
deal of information is lost and efforts should concentrate in the improvement
of high resolution theoretical data. Such an improvement must be based on a
quantitative validation process by comparing theoretical data with stellar
observations. On this line, \cite{Peterson01,
  Peterson03} and \cite{Peterson05} have embarked into a project aimed at providing a
large theoretical database fully consistent with Space Telescope Imaging
Spectrograph (STIS) observations.

In this paper we present a quantitative comparison between observed mid-UV
features in the form of absorption line and continuum indices and those
computed from high resolution synthetic spectra. We start, in
Section~\ref{sec:theory}, by investigating the effects of stellar parameters
and instrumental resolution on spectroscopic indices from a purely theoretical
point of view. In Section~\ref{sec:emp_vs_th} we compare theoretical and
observed indices and discuss the extent to which empirical indices are
reproduced by theory. Section~\ref{sec:calib} is devoted to provide the
equations needed to transform theoretical (discrepant) indices to the standard
(IUE) system. This represents the first attempt to derive synthetic indices from
high resolution mid-UV spectra.


\section{Theoretical UV indices}
\label{sec:theory}
Following the definitions given in Table~\ref{tab:indices}, we
carried out the calculation of a set of mid-ultraviolet indices of the
synthetic stellar SEDs of the UVBLUE
library\footnote{http://www.bo.astro.it/$\sim$eps/uvblue/uvblue.html and
  http://www.inaoep.mx/$\sim$modelos/uvblue/uvblue.html.}, described in
\citet[hereafter Paper~I]{Rodriguez05}. The indices measure either the
strength of absorption features or the slope of the continuum. It is important to mention
that most of the definitions listed in Table~\ref{tab:indices} correspond to those provided by 
\citealt{Fanelli90}. Two continuum indices (S2850 and S2850L) are analogous to the index defined by
LFB00. The band passes for these two indices and for 2110/2570 are those of Bressan 
(private communication). In column 5 of Table~\ref{tab:indices} we list the atomic 
species that most contribute to the absorption in the central band of each line 
index. The list has been assembled from a visual inspection of
the relevant absorption features present in the theoretical spectrum with $T_\mathrm{eff}$/$\log{g}$/[M/H]=5000/4.0/0.0.

The general procedure followed by \citet{Fanelli90} to define the wavelength
sequence and measure spectroscopic indices was carried out in a way similar to
that described by \citet{OConnell73}. For the calculations presented here,
UVBLUE has been degraded to match IUE nominal resolution (FWHM=6~\AA). The
line indices are the ratio of the integrated flux $F_i(\lambda)$ in the index
passband to the integrated continuum flux $F_c(\lambda)$, obtained by
linear interpolation of the mean flux of the two side bands; the result is
transformed to magnitudes:

\begin{equation}
I = -2.5 \times \log_{10} {\frac{\int_{\lambda_1}^{\lambda_2} {F_i(\lambda)
      \, d\lambda} } {\int_{\lambda_1}^{\lambda_2} {F_c(\lambda) \,
      d\lambda}}} \, ,
\end{equation}
where $\lambda_1$ and $\lambda_2$ are the limits of the index passband.

The continuum indices are given by the ratio of the mean flux of the blue
passband, $\overline{F}_b(\lambda)$, to the mean flux of the red one, $\overline{F}_r(\lambda)$, transformed to magnitudes:

\begin{equation}
I = -2.5 \times \log_{10} {\frac{\overline{F}_b(\lambda)}{\overline{F}_r(\lambda)}} \, .
\end{equation}

In the mid-UV, the spectra of intermediate and late-type stars are strongly
affected by line blanketing and therefore it is hard to find spectral regions
free from features. For this reason the spectral bands are selected such as to
optimize each index for a temperature interval in which the feature is
maximum. 

\begin{figure}[!h]
\begin{center}
\begin{tabular}{llll}
\resizebox{4cm}{!}{\includegraphics{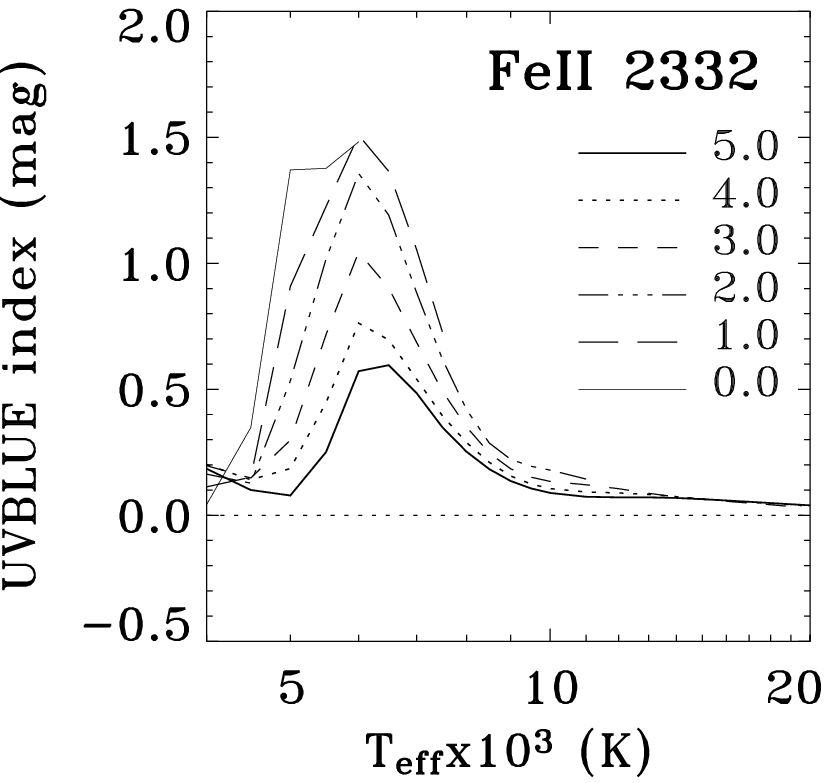}}   &
\resizebox{4cm}{!}{\includegraphics{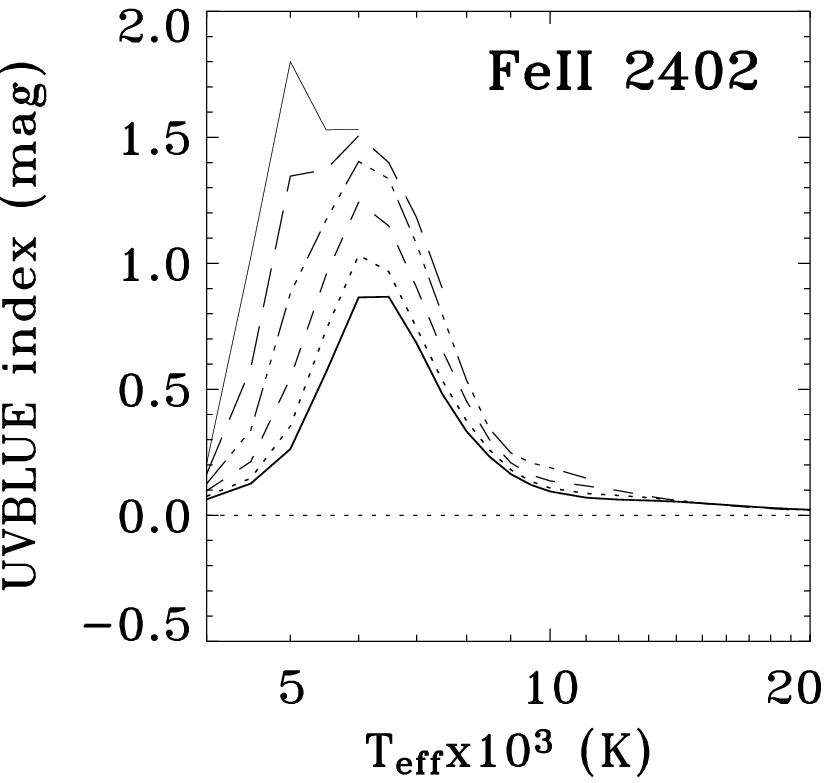}}   &
\resizebox{4cm}{!}{\includegraphics{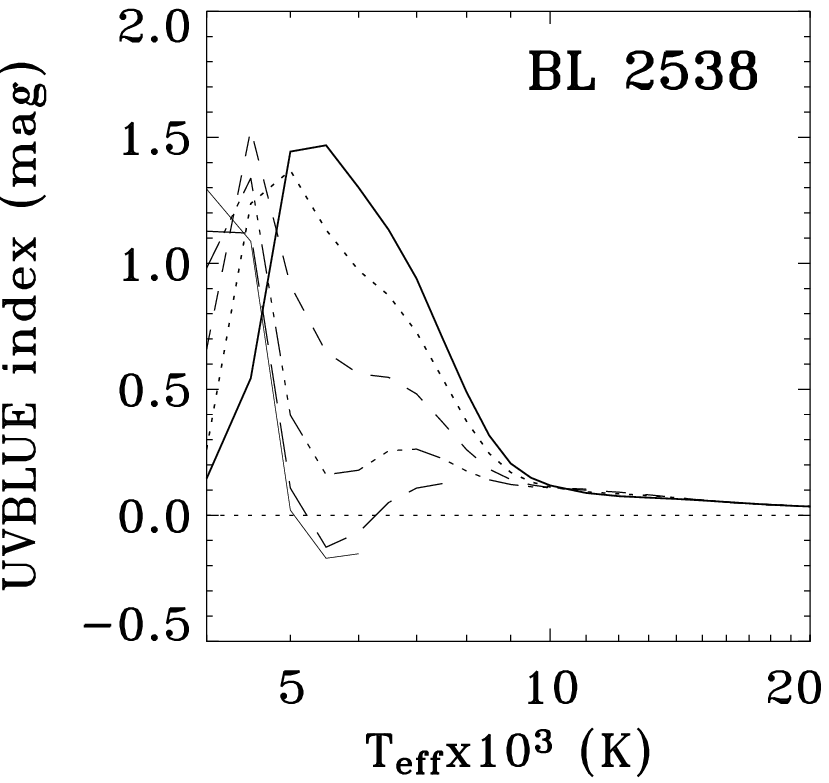}}     &
\resizebox{4cm}{!}{\includegraphics{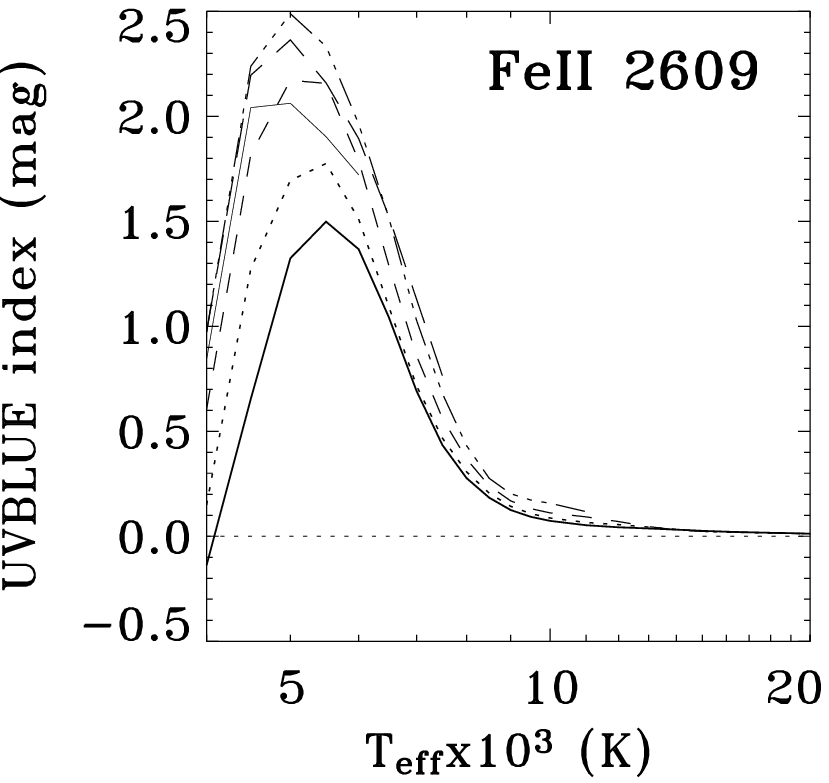}}   \\
\resizebox{4cm}{!}{\includegraphics{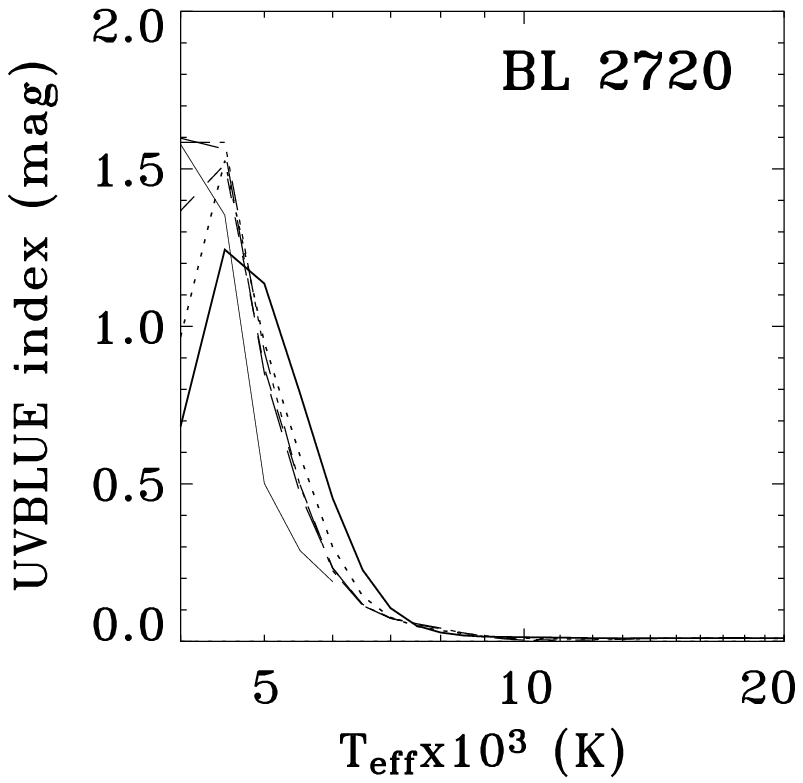}}   &
\resizebox{4cm}{!}{\includegraphics{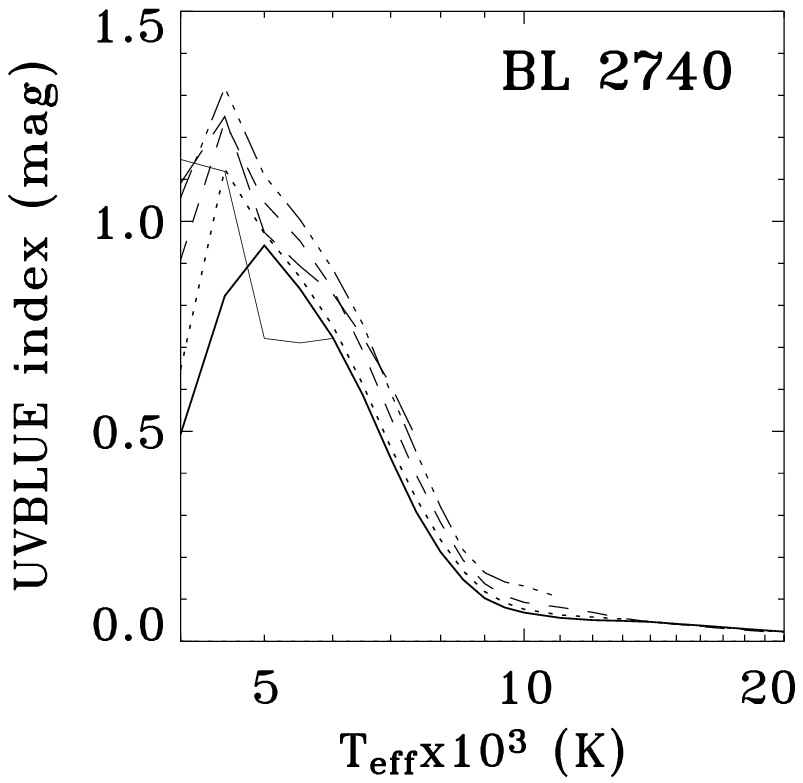}}     &
\resizebox{4cm}{!}{\includegraphics{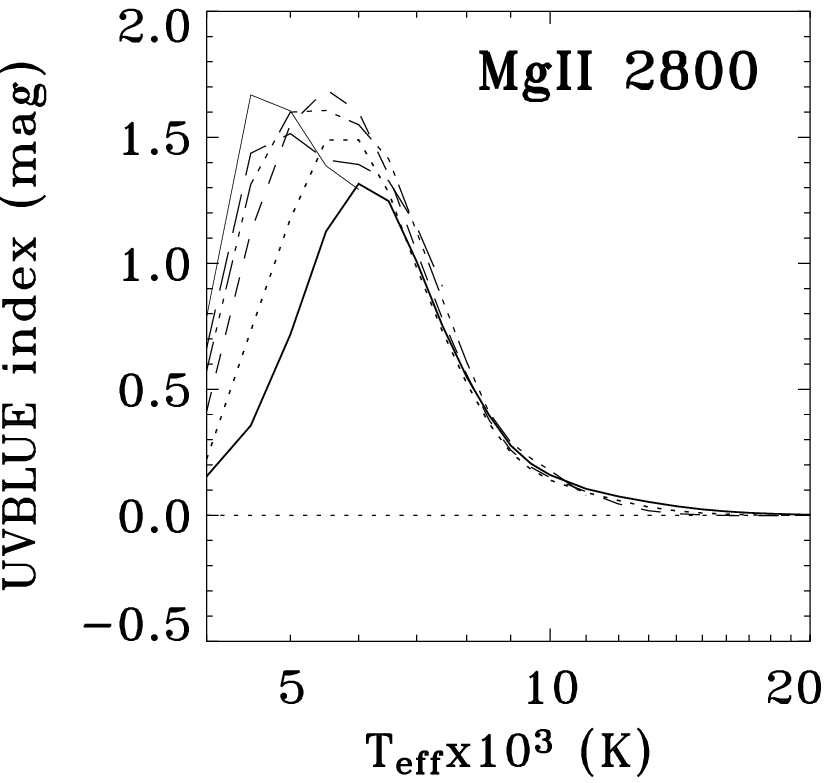}}   &
\resizebox{4cm}{!}{\includegraphics{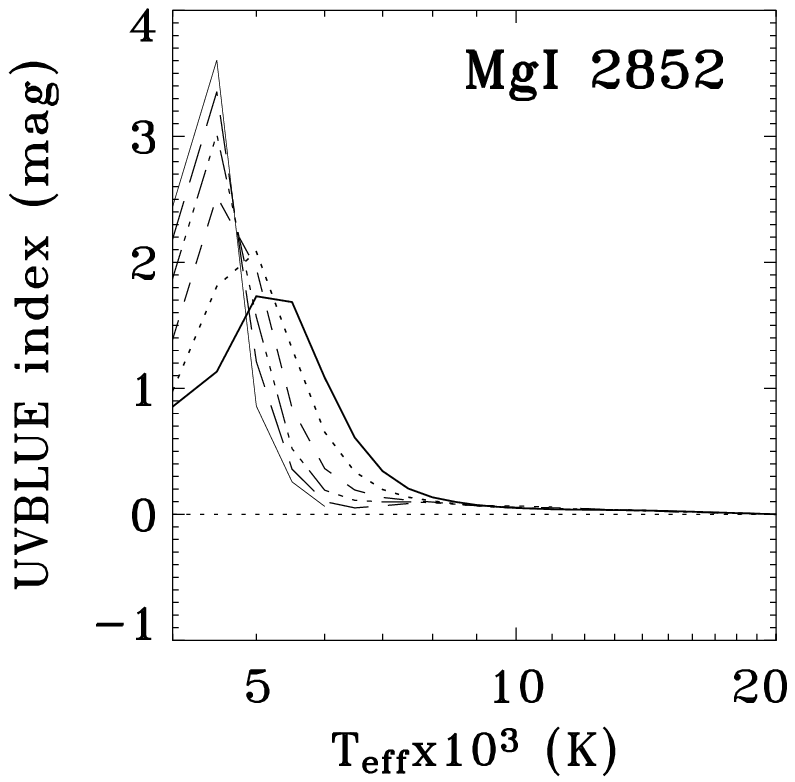}}    \\
\resizebox{4cm}{!}{\includegraphics{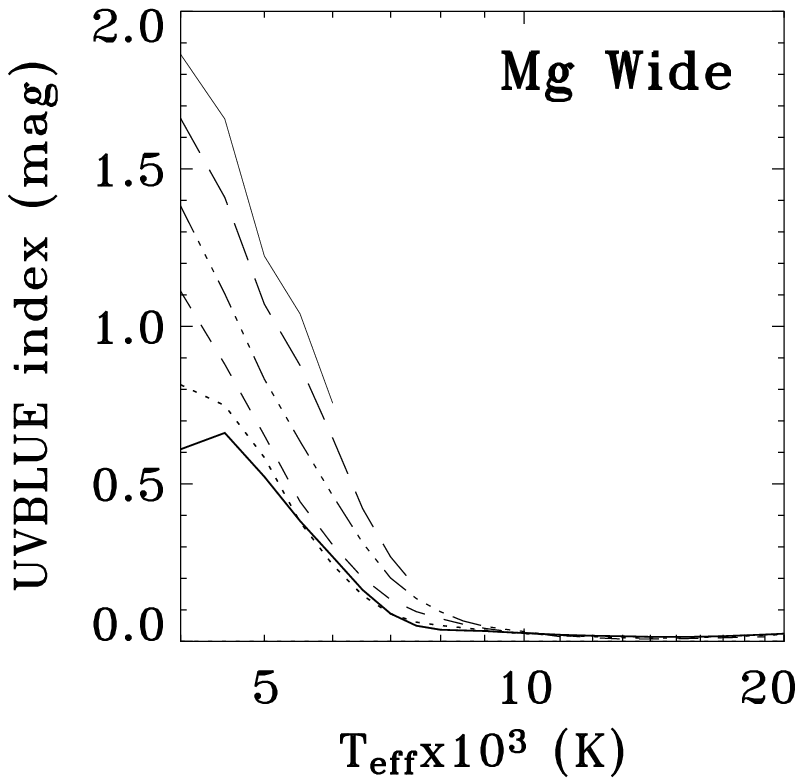}}     &
\resizebox{4cm}{!}{\includegraphics{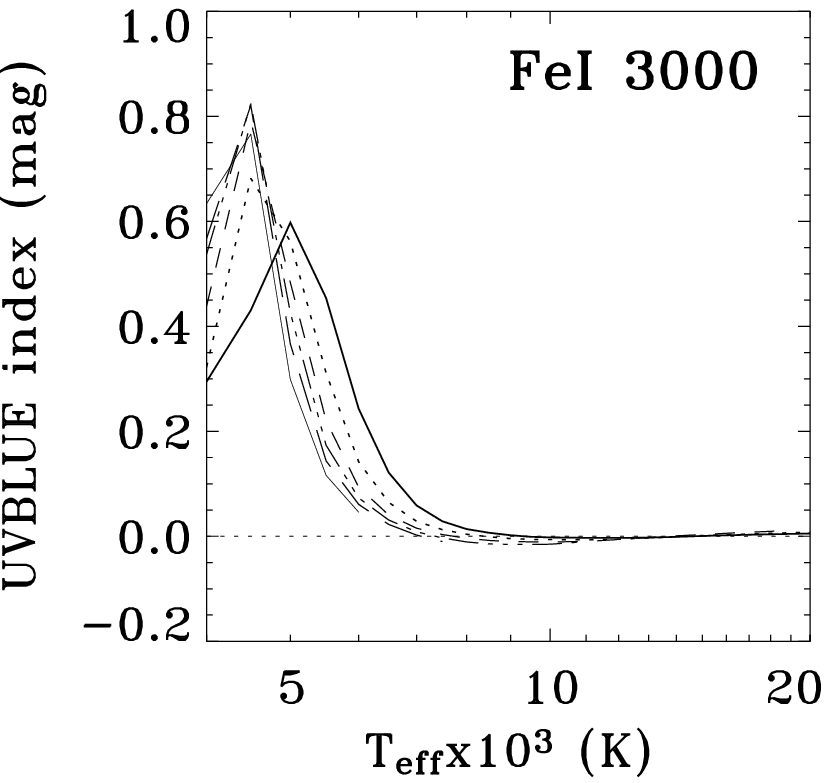}}    &
\resizebox{4cm}{!}{\includegraphics{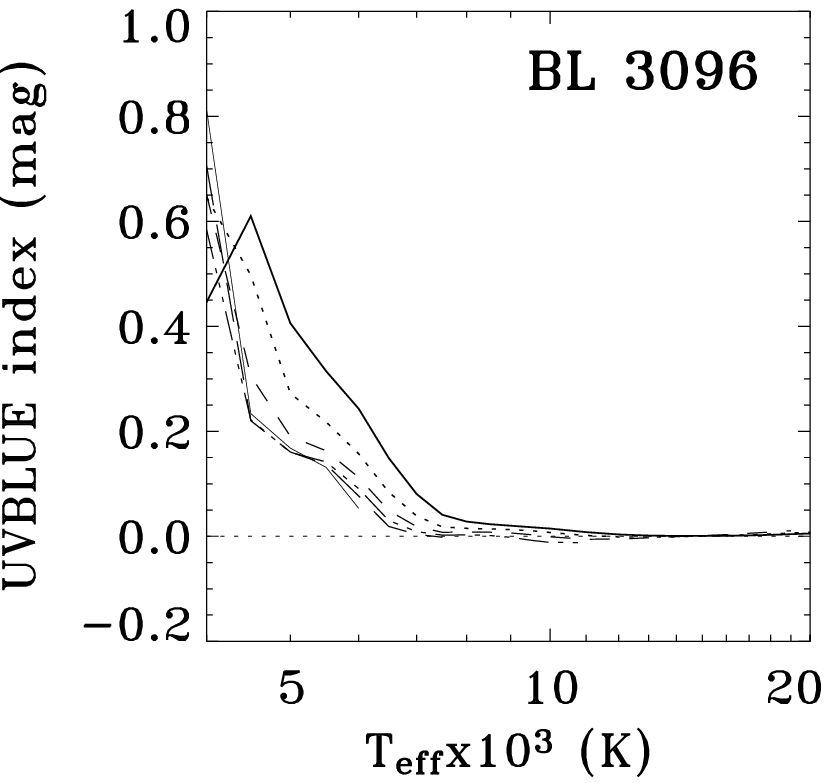}}     &
\resizebox{4cm}{!}{\includegraphics{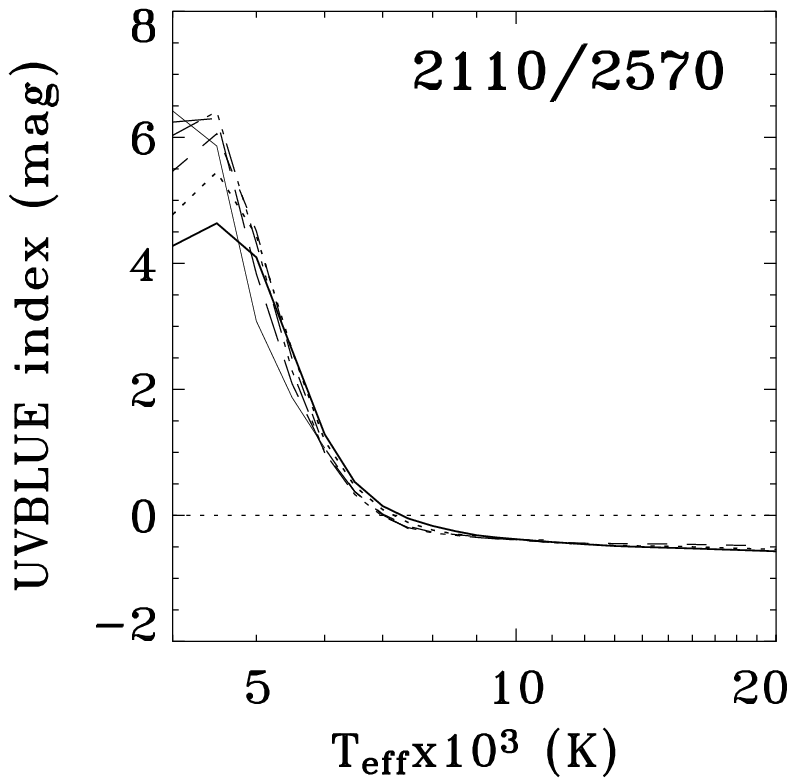}} \\
\resizebox{4cm}{!}{\includegraphics{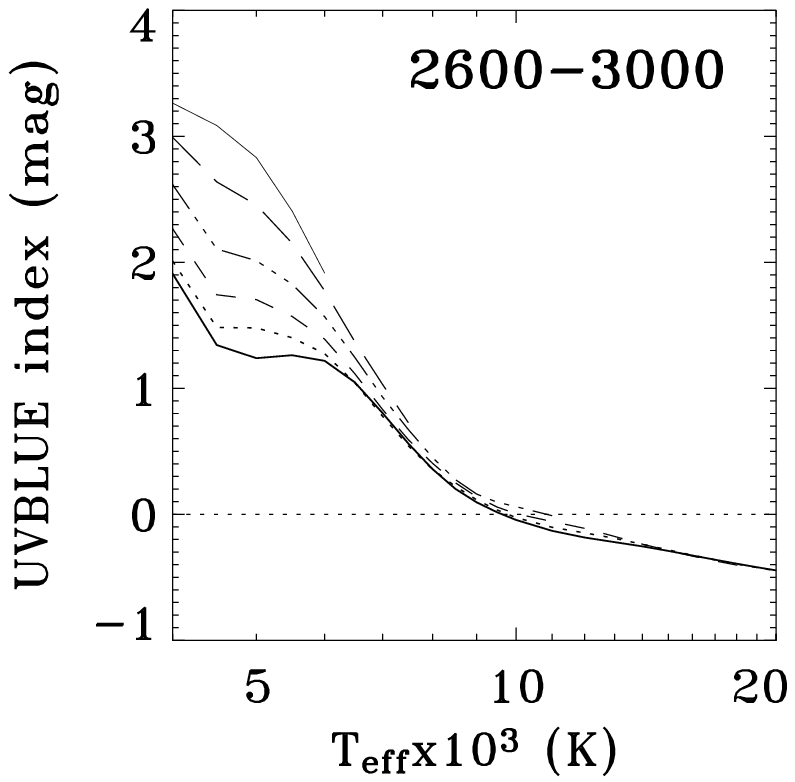}} &
\resizebox{4cm}{!}{\includegraphics{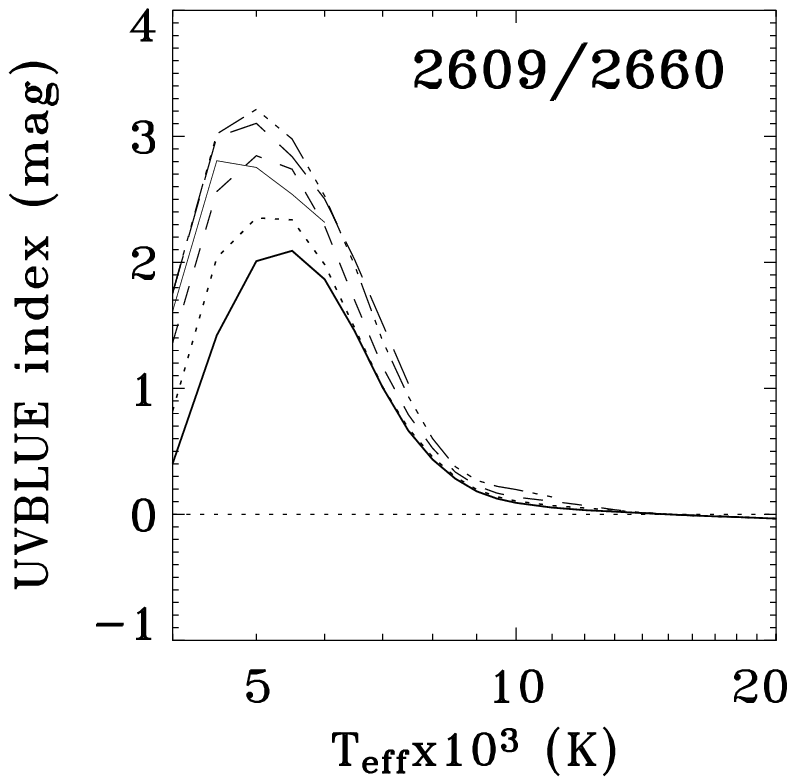}} &
\resizebox{4cm}{!}{\includegraphics{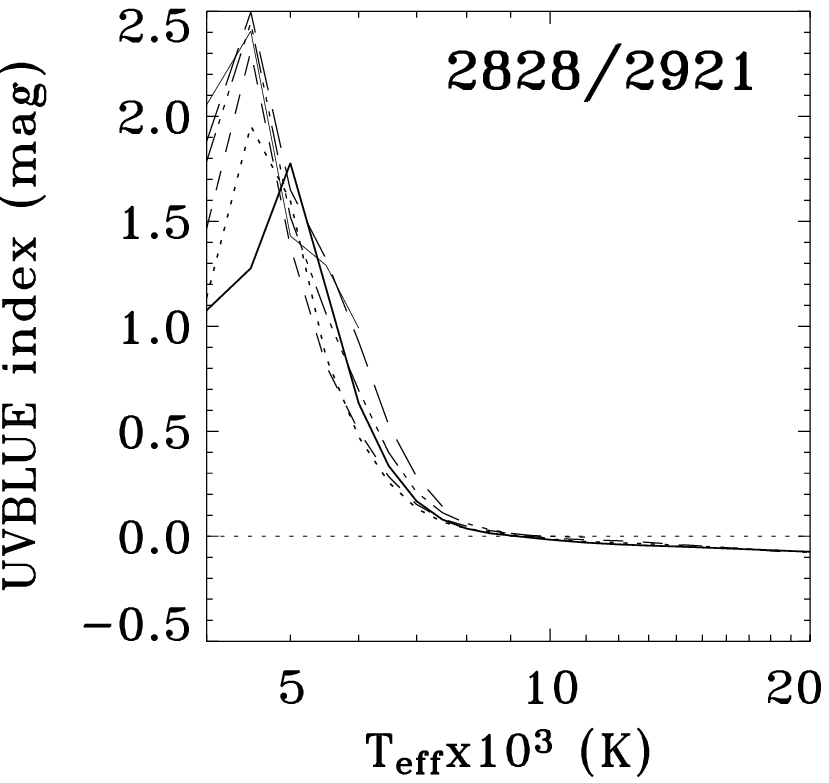}} &
\resizebox{4cm}{!}{\includegraphics{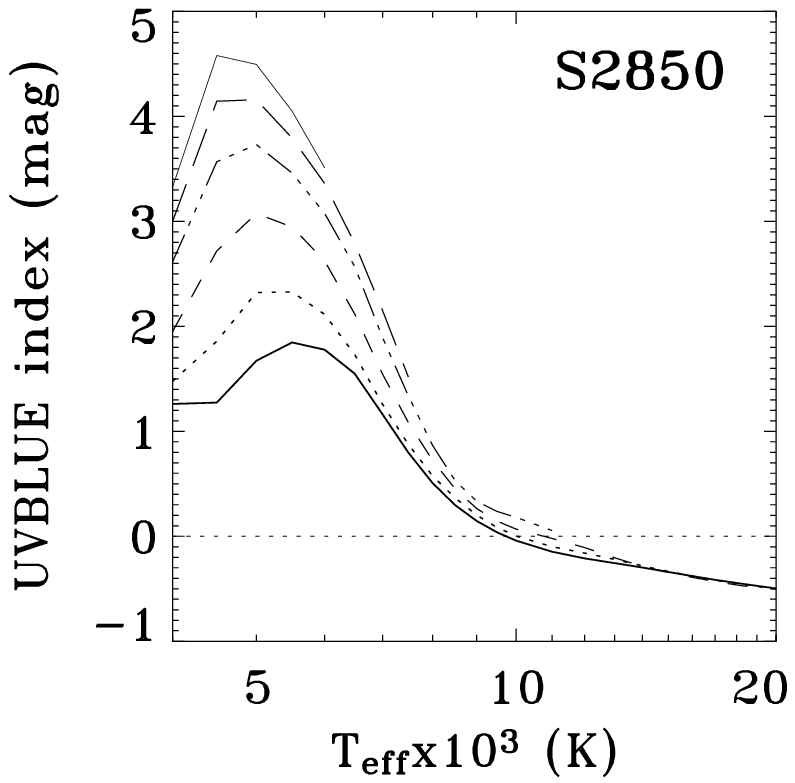}}      \\
\resizebox{4cm}{!}{\includegraphics{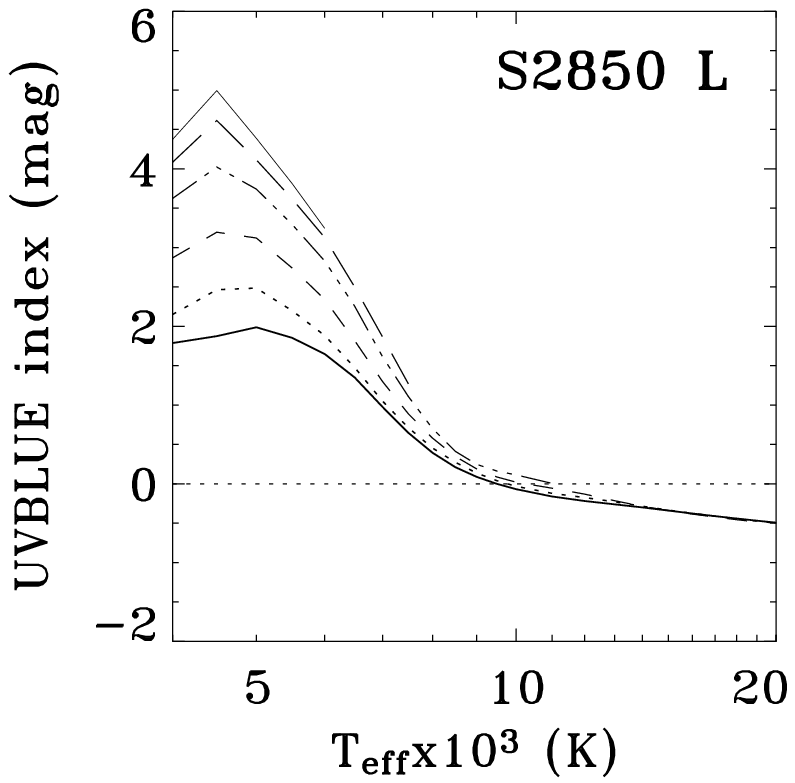}}    \\
\end{tabular}
\caption{Synthetic UV spectral indices as a function of effective temperature 
and surface gravity. All indices were computed for solar-metallicity and were properly degraded to match the IUE resolution. Different line-types indicate different gravity data sets as indicated in the top left panel.  \label{fig:idxteffgrav}}
\end{center}
\end{figure}

The homogeneous coverage in stellar parameters, in particular the chemical
composition, allows the investigation of their effects on each index.
The results are partially presented in Table~\ref{tab:uvblueindices} where we list some synthetic indices
(line indices) for a subset of atmospheric parameters in UVBLUE, identified in columns 1-3. 
The full grid of indices is avaliable in electronic form. In Figure~\ref{fig:idxteffgrav} we plot the index
vs. effective temperature ($T_\mathrm{eff}$) and surface gravity ($\log{g}$)
for synthetic spectra of solar metallicity. The reference labels inscribed in
the upper left panel stand for different gravities which range from
$\log{g}=0.0$ to 5.0~dex. We would like to stress at
this point that the approach in this section is purely theoretical and not
meant to analyze the compatibility between theoretical and empirical
indices. 

Some general remarks can be drawn just on the basis of a quick-look analysis 
of the different plots of Fig.~\ref{fig:idxteffgrav}, simply relying on prime physics
principles. 
In all cases it is evident that indices probe a characteristic temperature range 
at which the corresponding absorption feature (or pseudo-continuum break) is maximum.
As expected, metal indices for higher ionization states generally peak at warmer effective 
temperature: this is because stellar layers at the appropriate thermodynamical 
conditions to produce the feature lie at a shallower optical depth with increasing stellar 
effective temperature.
This is well evident, for instance, looking at the Mg~\textsc{ii}~2800 and Mg~\textsc{i}~2852 indices
or comparing the Fe~\textsc{ii}~2402 feature with the Fe~\textsc{i}~3000 one. In both cases, the 
higher ionization state is the prevailing one among F stars while the neutral
metal becomes more important for the G spectral type.

As for gravity effects, one has to consider that the relative partition of a
given elemental species between, let us say, first ($X_{\sc II}$) and neutral
($X_{\sc I}$) ionization states scales according to the Saha equation
\citep[e.g.][]{Mihalas78} as
\begin{equation}
{X_{\sc II}\over X_{\sc I}} \propto {{T^{3/2} e^{-1/kT}} \over P_e}.
\label{eq:saha}
\end{equation}
with $P_e$ being the electronic pressure of the stellar plasma.

\begin{figure*}[ht]
\begin{center}
\includegraphics[width=\hsize]{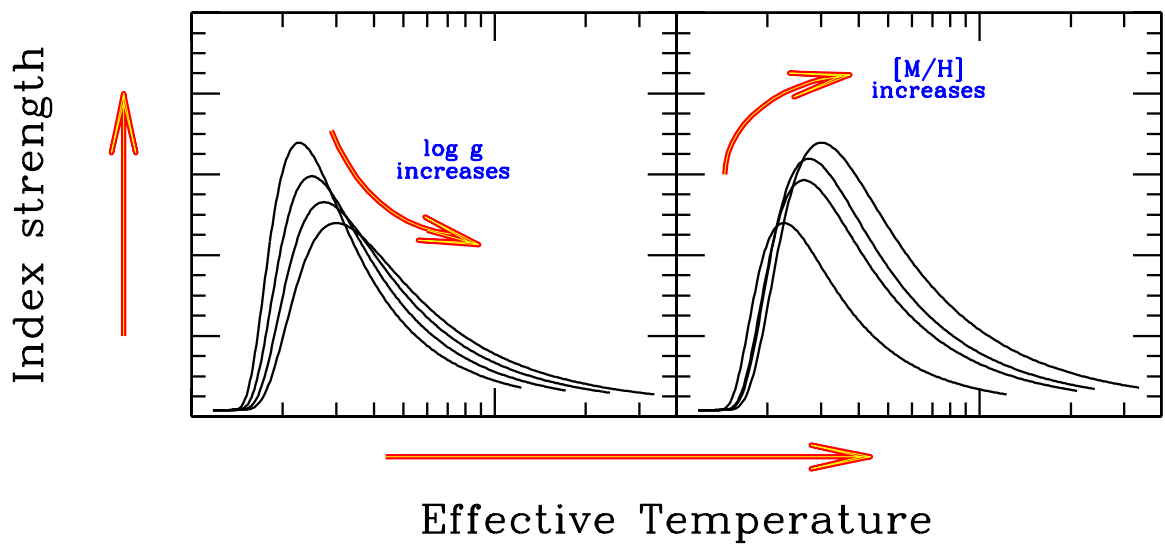}
\caption{The expected effect on metal indices by increasing electronic pressure ($P_e$) in 
stellar atmosphere. This effect can be induced either by increasing surface gravity $\log g$
(that is by ``packing'' atoms more efficiently) or by increasing metal abundance (thus
increasing the main $e^-$ donors to the plasma). Note the different trend of the index strength
depending whether gravity or metallicity are acting. In the first case (left panel) we expect a weaker 
index for dwarf stars compared to giants due to the wiping effect of damping broadening in the
spectral features with increasing gravity. Conversely (right panel), the index tends to be stronger if we
increase the corresponding elemental abundance.}
\label{fig:trend}
\end{center}
\end{figure*}

Therefore, with increasing $\log g$ (that is by ``packing'' atoms more efficiently)
one makes easier the ionization process allowing metals to more efficiently feed $e^-$ to the plasma.
As a consequence, the correspondingly higher value of $P_e$ leads any given equilibrium partition
$X_{\sc II}/X_{\sc I}$ to shift to slightly higher temperatures. So, in Fig.~\ref{fig:idxteffgrav} a given 
spectral feature is seen to peak at earlier spectral types with increasing $\log g$ (see, for instance,
the nice trend of Fe~\textsc{ii}~2402 or Mg~\textsc{ii}~2800 in the figure). In addition, the broader
damping wings of high-gravity spectral features mimic the wiping effect of a lower spectral
resolution (see Sec.~\ref{sec:resolution}) thus favoring weaker index values for dwarf stars
compared to giants. These qualitative arguments are summarized in Fig.~\ref{fig:trend}.

Based on a more careful inspection of Fig.~\ref{fig:idxteffgrav}, we can give
below further comments on individual indices:

{\it Fe~\textsc{ii}~2332}: the absorption in the central band of this index is
mainly due to Fe~\textsc{ii} and Fe~\textsc{i} lines, although several other
elements add significant contribution, as in all other indices that we present
here. This index has its maximum for all gravities at
about $T_\mathrm{eff}=6000$~K. It displays a large variation with surface
gravity, ranging from 0.5~mag for high gravity stars to 1.5~mag for supergiant
spectra.

{\it Fe~\textsc{ii}~2402}: feature produced by multiplets of Fe~\textsc{ii}. 
This index is very sensitive to gravity. At its maximum, at
$T_\mathrm{eff} = 6000$~K, the index doubles its value from about 0.8 to
1.5~mag when the gravity decreases from $\log{g} = 5.0$ to 1.0. All gravities
peak at about the same temperature (6000~K).

{\it BL~2538}: the blend is produced by mainly Fe, Mg, Cr, and Ni absorption. 
Among line indices, BL~2538 shows the highest
sensitivity to surface gravity. It peaks (index~$\approx 1.5$~mag) at the
lowest temperature (4000~K) for $\log{g} = 5.0$. For $\log{g} = 0.0$ the index
drastically drops to negative values. Note that, contrary to most indices, for
$T_\mathrm{eff}$ above 4500~K this index increases with increasing gravity.

{\it Fe~\textsc{ii}~2609}: the absorption in the index band is largely due to
Fe~\textsc{ii}, with contributions by Fe~\textsc{i} and Mn~\textsc{ii}. 
This index has its maximum at
about 5000~K, temperature at which it displays a strong sensitivity to gravity
with a variation from 1.5~mag, for $\log{g} = 5.0$, to 2.5~mag for $\log{g} =
2.0$. It peaks at about the same temperature for all gravities.

{\it BL~2720}: blend produced by neutral and ionized iron lines, with presence of
Cr~\textsc{i} and V~\textsc{ii}. Down to $T_\mathrm{eff}=4500$~K,
this index shows little variation with surface gravity, with the exception of
the outermost $\log{g}$ values. Similarly to BL~2538 for high gravity, this
index reaches its maximum at the lowest temperatures.

{\it BL~2740}: blend of iron and chromium lines. This index is similar to
BL~2720 in that it shows marginal dependence with gravity, it sharply
increases with decreasing $T_\mathrm{eff}$, and it peaks at the lower
temperature edge of UVBLUE.

{\it Mg~\textsc{ii}~2800}: this index includes the strongest feature in the
mid-UV interval, due to Mg~\textsc{ii}. It sharply increases shortward of
10000~K and peaks at about 6000~K for $\log{g}=5.0$ and 5000~K for
$\log{g}=0.0$. The index shows slight dependence on gravity since it varies
about 30\% along the full gravity interval. This dependence is restricted for
$T_\mathrm{eff} < 6000$~K.

{\it Mg~\textsc{i}~2852}: this index is produced by a strong Mg~\textsc{i}
absorption line. It displays an enhanced sensitivity to the surface gravity
with respect to Mg~\textsc{ii}~2800. Indices from high gravity spectra peak at
$T_\mathrm{eff}=6000$~K, while for the lowest gravity the maximum value is
reached at about 4500~K.

{\it Fe~\textsc{i}~3000}: the absorption in the index band is dominated by
Fe~\textsc{i}. It appears to be sensitive to $T_\mathrm{eff}$ in a short
range, $4000 \leq T_\mathrm{eff} \leq 8000$~K, and it is moderately dependent
on gravity.

{\it BL~3096}: the blend is produced by Fe~\textsc{i}, Ni~\textsc{i},
Mg~\textsc{i}, and Al~\textsc{i}. It displays a marginal sensitivity on
gravity for high gravity spectra. Giants and supergiants are not separated
among themselves.

{\it Mg Wide }: this index covers 600~\AA\ of the mid-UV SED. Its central band
contains those of BL~2720, BL~2740, Mg~\textsc{ii}~2800, and
Mg~\textsc{i}~2852. Is has been
envisaged as a useful indicator of the effective temperature of the dominant
stellar population in evolved systems. Defined by three 200~\AA-wide bands, this index
is expected to be suitable for the analysis of low resolution data of distant
galaxies collected by current large facilities \citep{Heap98,McCarthy04}. The curves
depicted in Fig.~\ref{fig:idxteffgrav} show that the index increases
monotonically for most gravities up to the low temperature edge; only for
$\log{g} = 5.0$ the index saturates at 4500~K (early K-type stars). Mg~Wide is
very sensitive to surface gravity with indices derived from lower gravity
spectra displaying the highest values.

The general behavior of continuum indices (spectral breaks and broad band
indices) is depicted below.

{\it 2110/2570}: this index primarily measures
the strong Mg~\textsc{i} magnesium break at 2512~\AA. It is interesting to note that
down to about 5000~K the index is insensitive to gravity. At lower
temperatures the effects of gravity are moderate. For synthetic fluxes with
$T_\mathrm{eff} > 7000$~K the index turn to negative values.

{\it 2609/2660}: as indicated by \citet{Fanelli90} the spectral break at
about 2600~\AA\ is dominated by iron opacity. In our synthetic spectra indices for all gravities peak at
about the same temperature (5500~K) with perhaps a slight shift towards lower
temperatures at lower gravities. The index shows a moderate sensitivity to
gravity, changing by 50\% over the entire $\log{g}$ interval. The break
vanishes for models of 10000~K.

{\it 2828/2921}: this break is among the most prominent features in the mid-UV
spectra in cool stars and evolved populations. It displays very low sensitivity to surface gravity
for model spectra of $T_\mathrm{eff} > 5000$~K. Below this temperature, which
marks the peak for high gravity fluxes ($\log{g} = 5.0$), the trends for
different gravities clearly separate.  At the lowest temperatures the change
of the index in the interval $\log{g}=0.0$--5.0 is about a factor of two.

{\it 2600-3000, S2850, and S2850L}: as can be seen from the bands listed in
Table~\ref{tab:indices}, these indices measure the slope between 2600 and
3000~\AA. The differences are the band widths and the centers of the bands of
the index 2600-3000. As expected, the overall behavior of the three continuum
indices is very similar: all turn from negative to positive values at
$T_\mathrm{eff} \sim 10000$~K, higher gravity model fluxes display larger indices, and all are
sensitive to gravity, with an enhanced sensitivity of the index S2850 that
varies almost a factor of three from $\log{g}=0.0$ to 5.0. 

We have described the general behavior of the 17 line and continuum indices
from a purely theoretical point of view. It is worth to remark that the 
above results are affected by the theoretical limitations of model
atmospheres, in particular, the limitations inherent to models constructed
under the classical approximations, as is our case. In Paper~I we have
commented on the UVBLUE caveats and the reader is referred to that paper for
more details. The discussion of the impact of UVBLUE limitations on synthetic
indices is deferred to Sec.~\ref{sec:emp_vs_th}.
Nevertheless, we can anticipate that the overall trends of index values
with $T_\mathrm{eff}$ and surface gravity are similar to those depicted by
indices computed from empirical data \citep{Fanelli92}. Note
that we have restricted the analysis to line indices
redward of 2300~\AA\ since both, theoretical and observed spectra, decrease in
quality for intermediate and late stellar types. In what follows we describe the effects of 
chemical composition, which, in the context of stellar populations analyses,
might be more important than those of gravity.

\subsection{Effects of metallicity}
The uniform metallicity coverage of UVBLUE (from [M/H]=$-$2.0 to $+0.5$) allows 
the investigation, at least from an exploratory point of
view, of the effects of chemical composition on spectroscopic indices. Such an
investigation is not feasible with current empirical databases since the vast
majority of objects have solar or nearly solar metallicity (see Fig.~8 of
Paper~I).

\begin{figure}[t]
\begin{center}
\begin{tabular}{llll}
\resizebox{4cm}{!}{\includegraphics{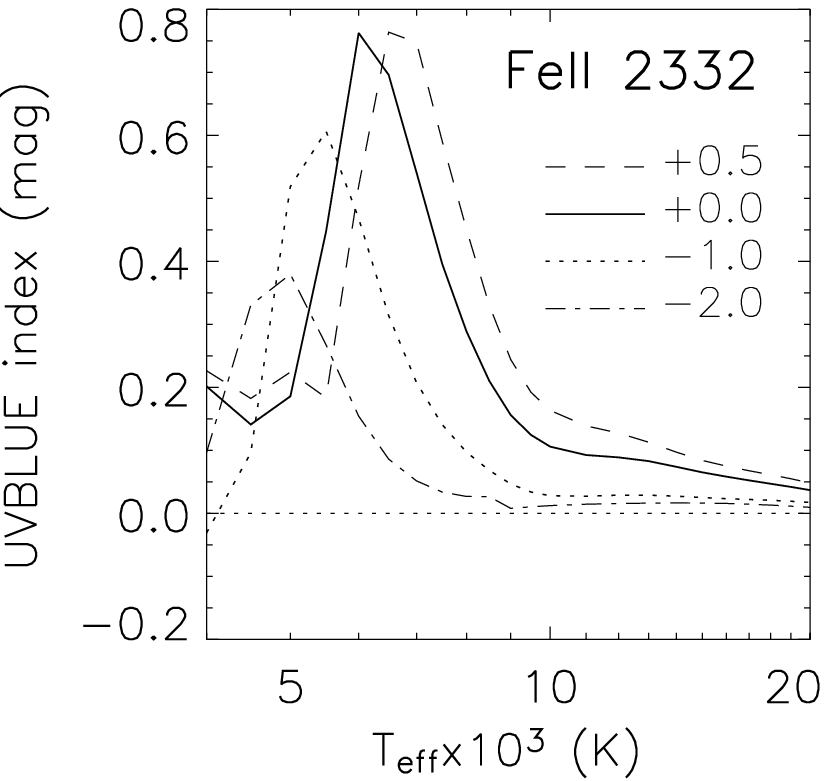}}   &
\resizebox{4cm}{!}{\includegraphics{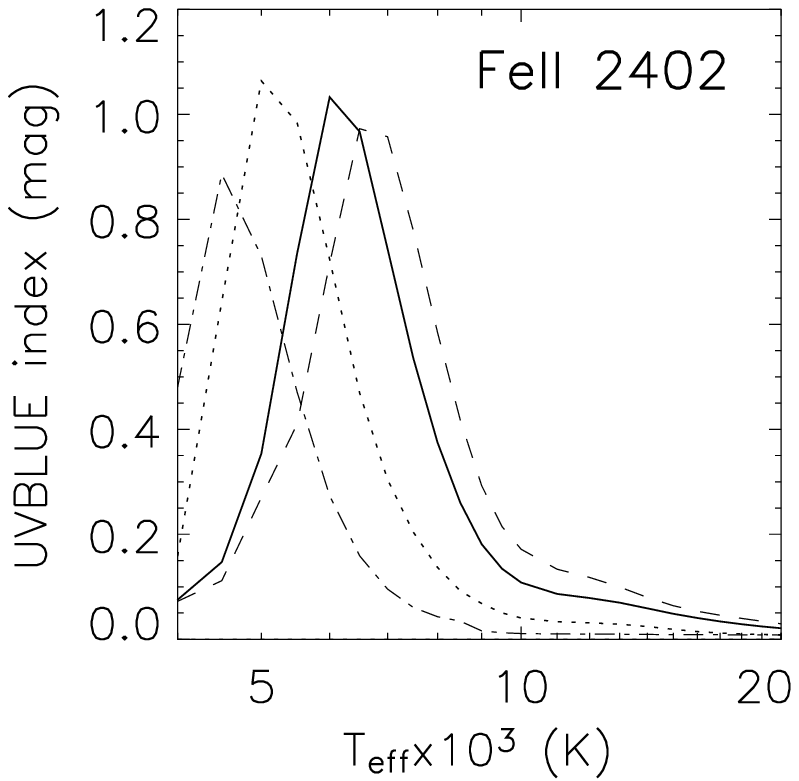}}   &
\resizebox{4cm}{!}{\includegraphics{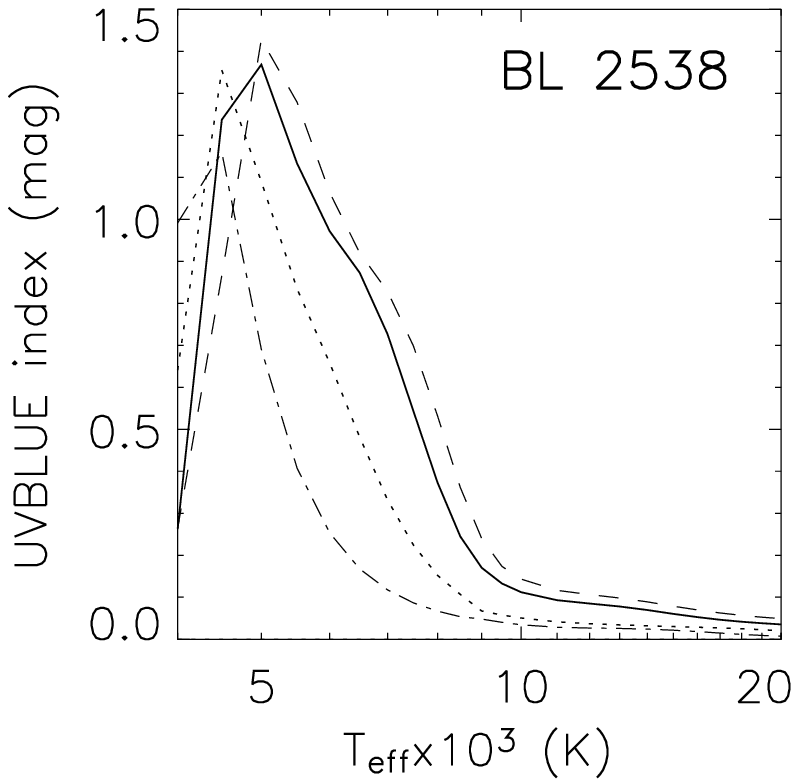}}     &
\resizebox{4cm}{!}{\includegraphics{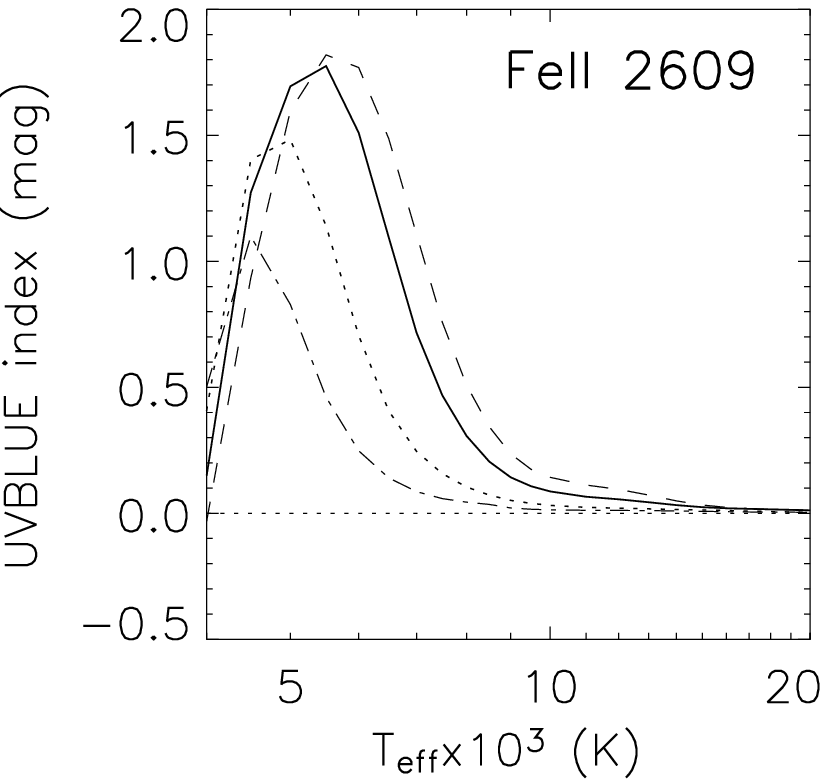}}   \\
\resizebox{4cm}{!}{\includegraphics{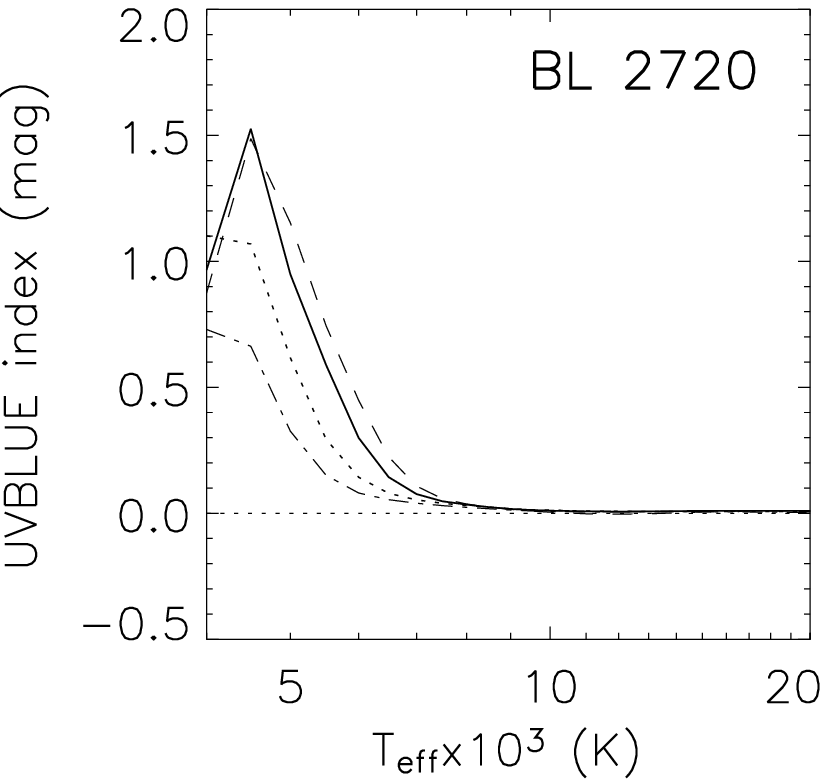}}   &
\resizebox{4cm}{!}{\includegraphics{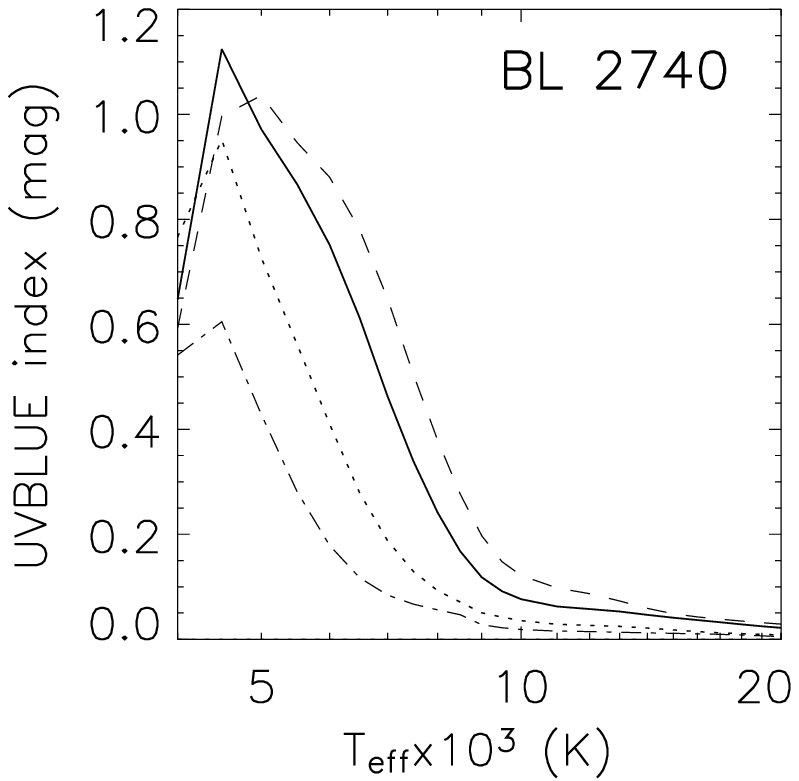}}     &
\resizebox{4cm}{!}{\includegraphics{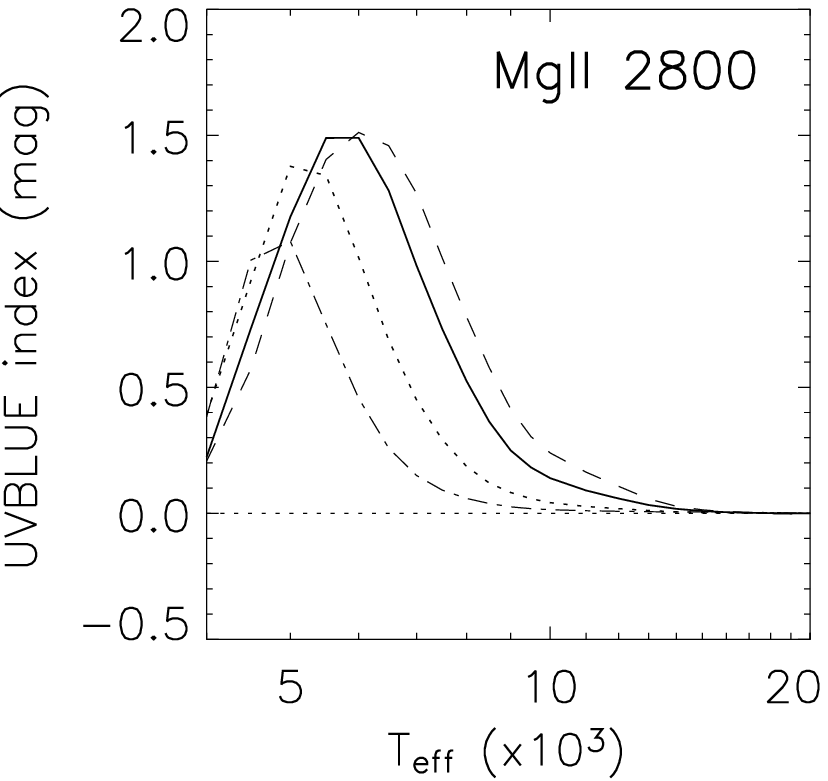}}   &
\resizebox{4cm}{!}{\includegraphics{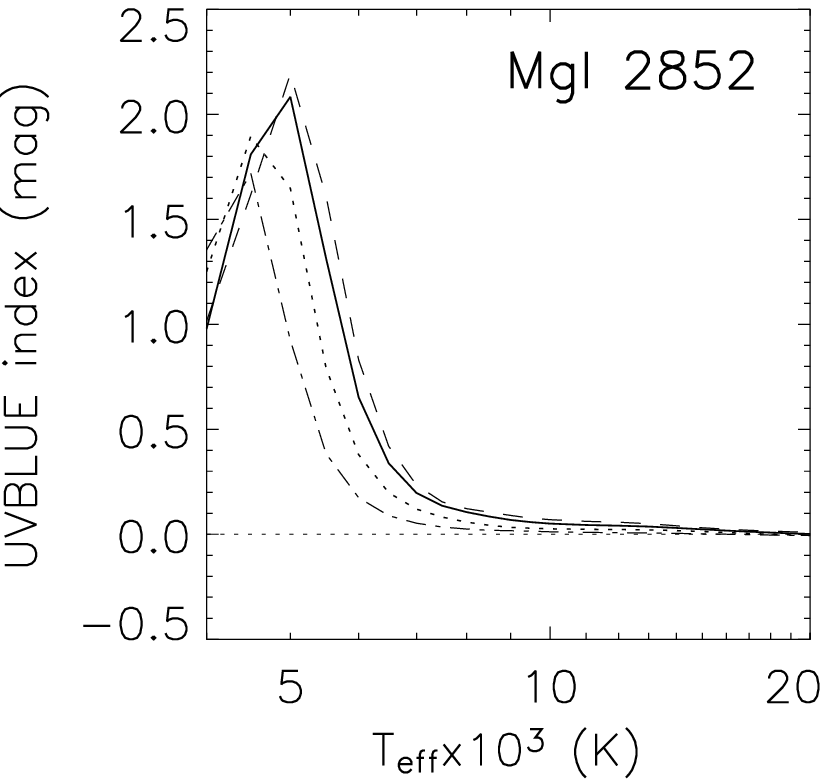}}    \\
\resizebox{4cm}{!}{\includegraphics{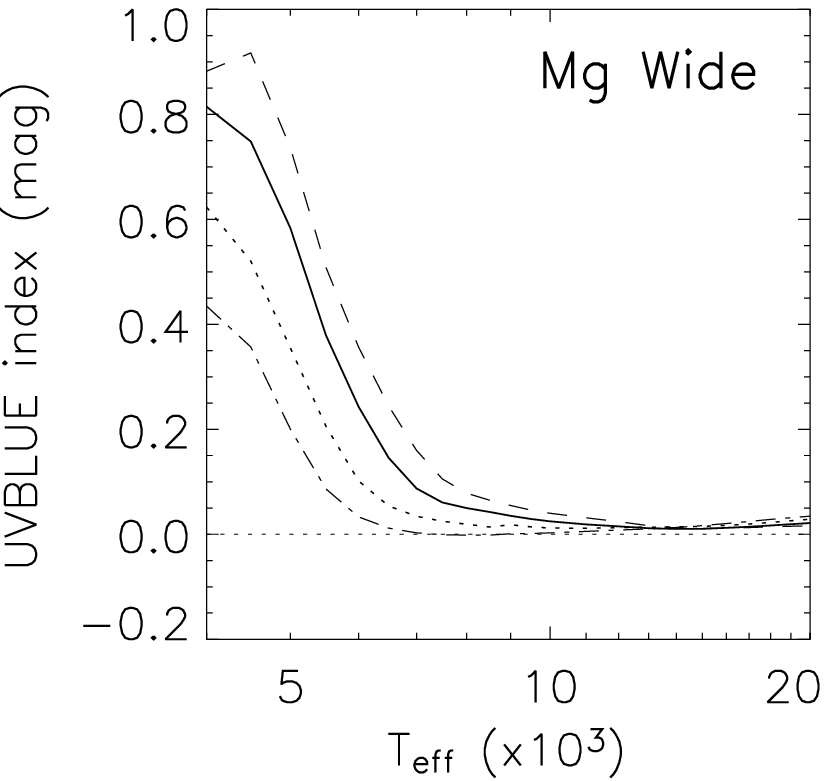}}     &
\resizebox{4cm}{!}{\includegraphics{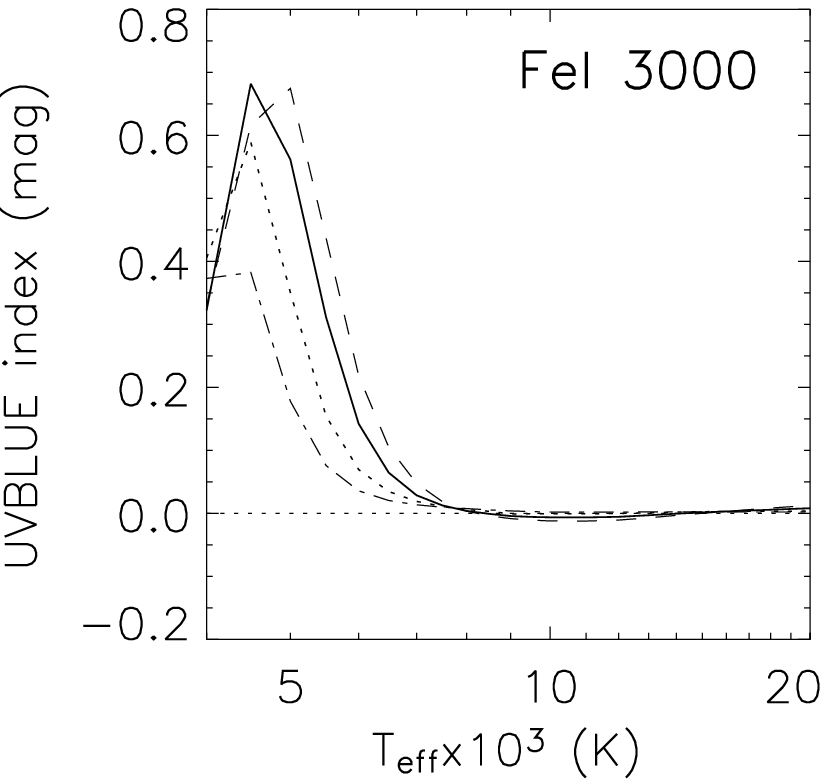}}    &
\resizebox{4cm}{!}{\includegraphics{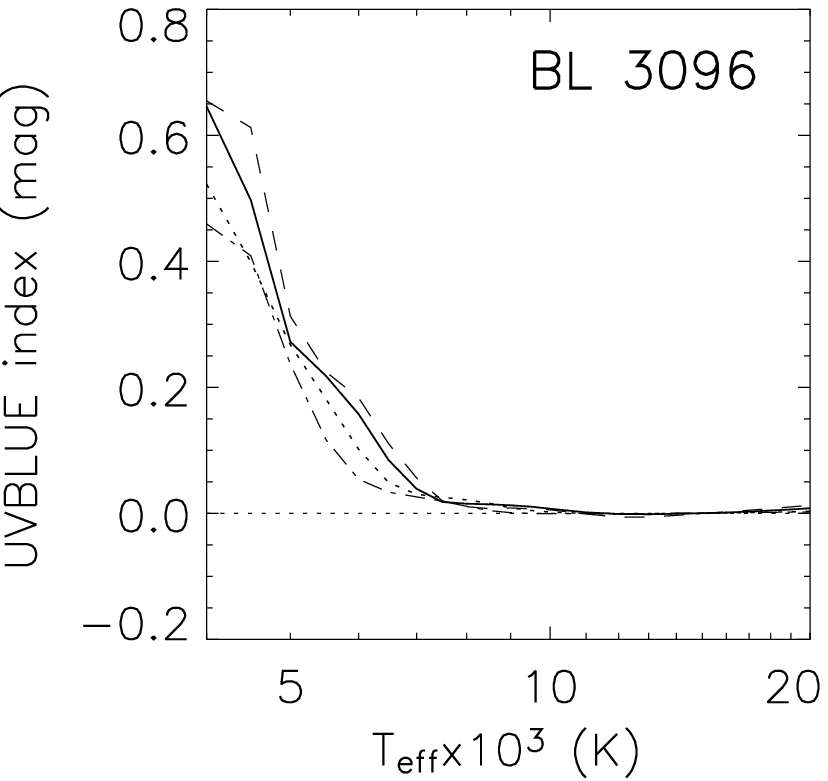}}     &
\resizebox{4cm}{!}{\includegraphics{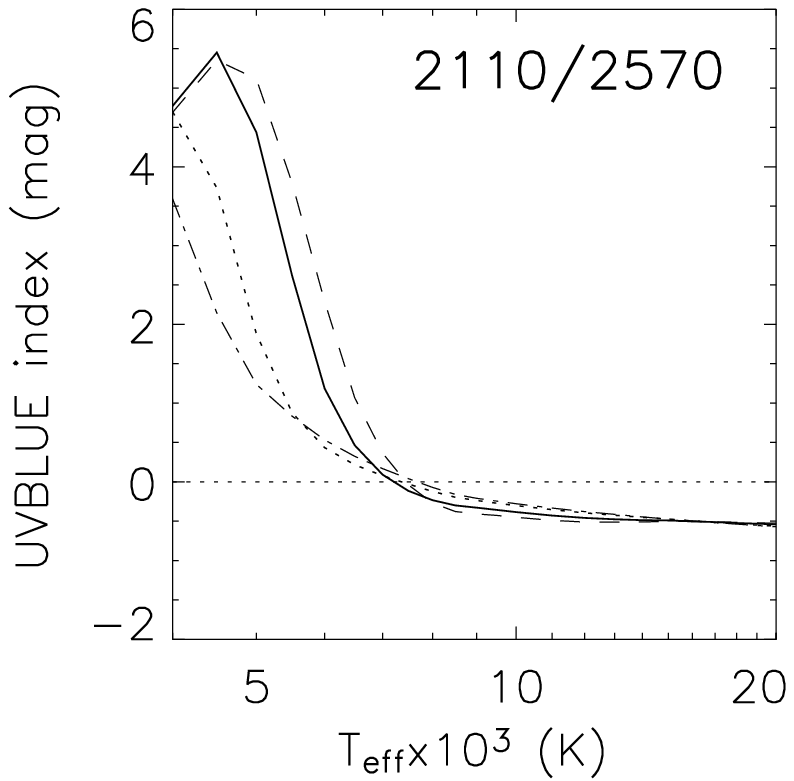}} \\
\resizebox{4cm}{!}{\includegraphics{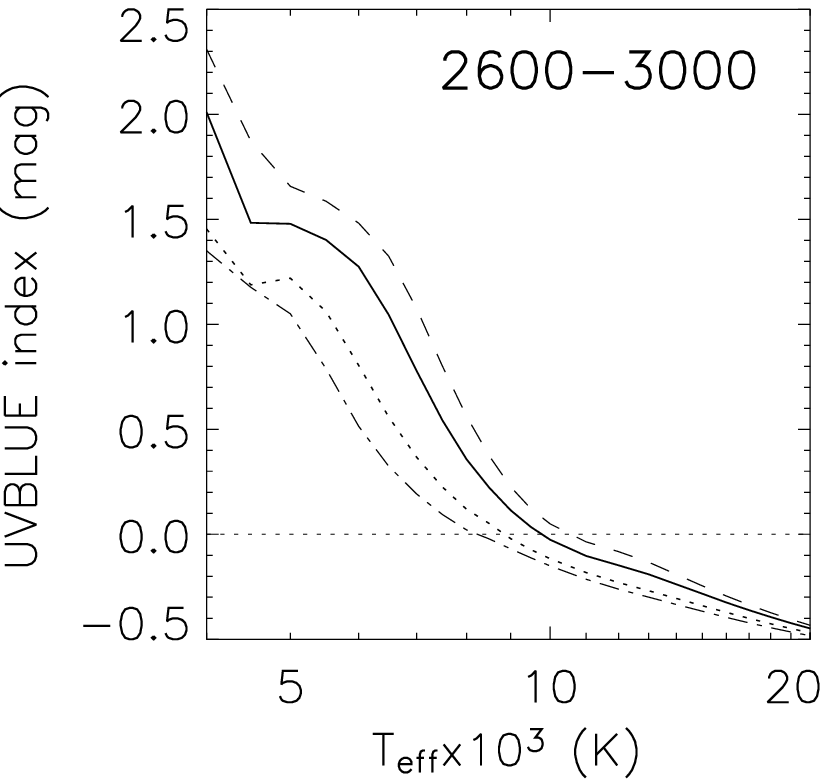}} &
\resizebox{4cm}{!}{\includegraphics{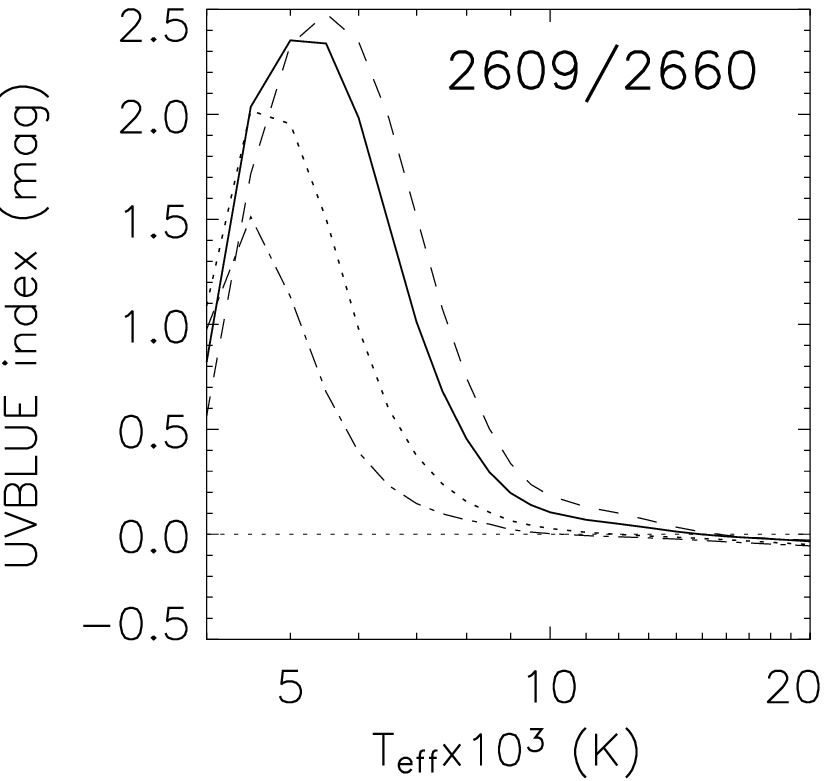}} &
\resizebox{4cm}{!}{\includegraphics{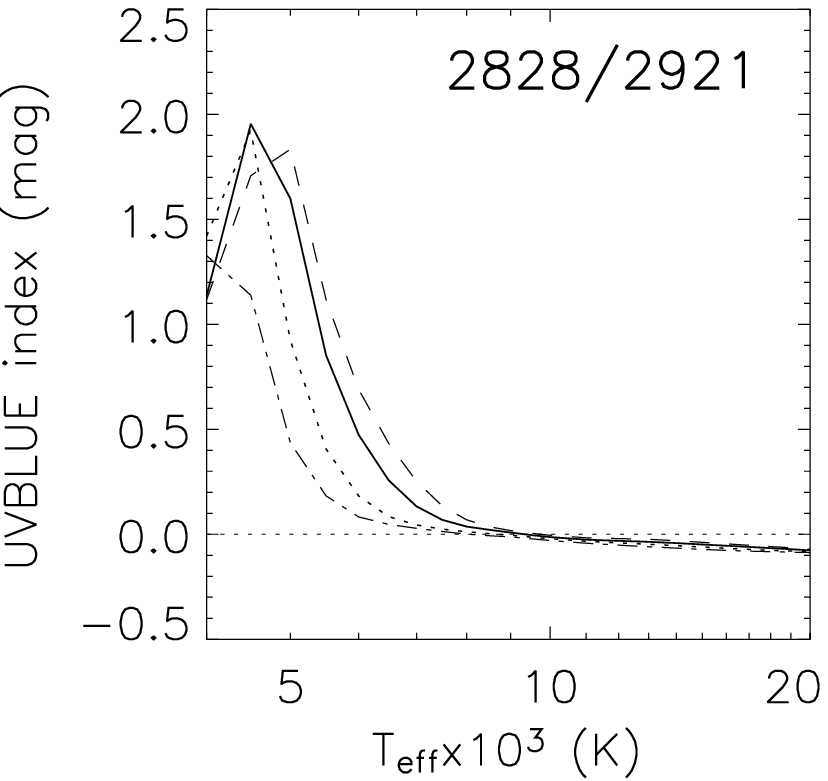}} &
\resizebox{4cm}{!}{\includegraphics{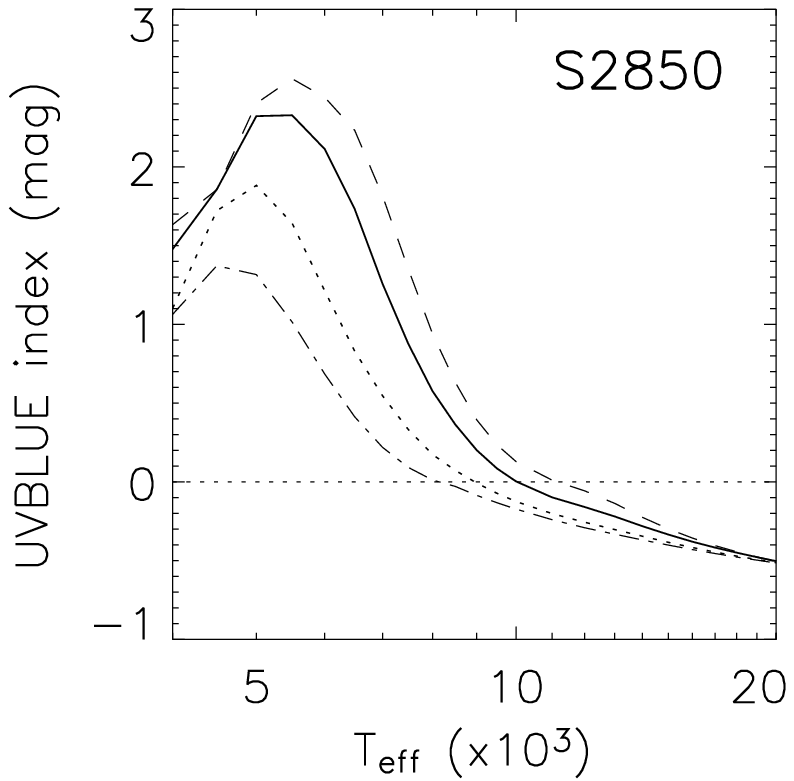}}      \\
\resizebox{4cm}{!}{\includegraphics{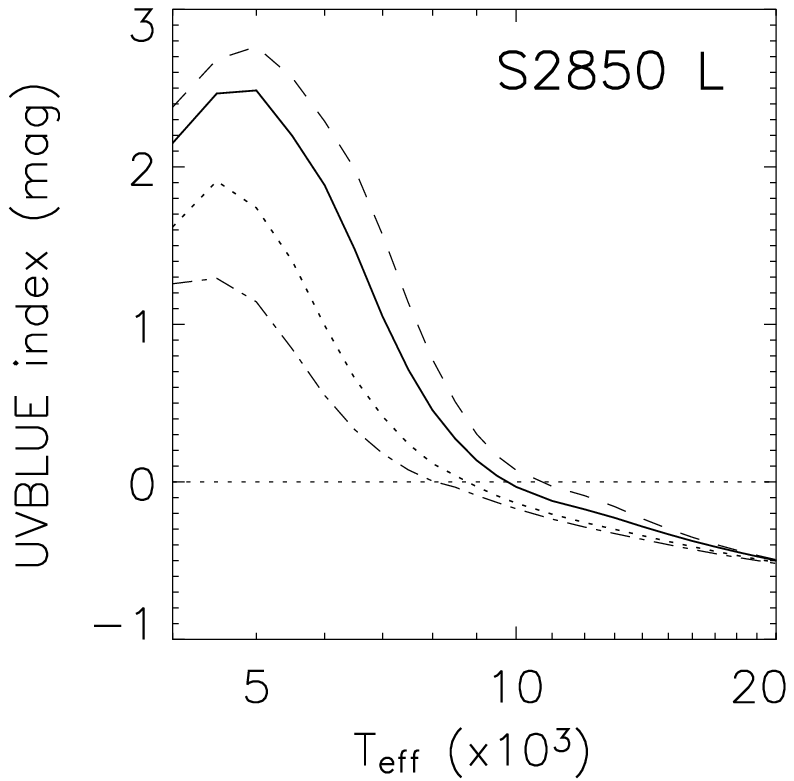}}    \\
\end{tabular}
\caption{Effects of chemical composition and effective temperature for spectra
  with $\log{g}$=4~dex. Different line-types represent, as labeled in the first panel, four metallicities ranging from super solar ([M/H]=+0.5) to subsolar ([M/H]=$-2.0$). \label{fig:idxteffmet}}
\end{center}
\end{figure}

In Fig.~\ref{fig:idxteffmet} we show the trends of the mid-UV spectral
indices as a function of the effective temperature and metallicity for a fixed
surface gravity of $\log{g}=4.0$~dex.
Different line types correspond to different metallicities as denoted in 
the labels in the upper left panel in the figure. 
Like for gravity effects, a similar trend on spectrophotometric indices 
has also to be expected by increasing stellar metallicity.
In this case, $P_e$ would increase in eq.~(\ref{eq:saha}) merely as a consequence of a 
higher value of $Z$. In addition, in case of thin absorption features, the corresponding 
index strength would increase reaching a stronger peak for higher values of metal abundance.
This is nicely shown in Fig.~\ref{fig:idxteffmet}, specially for the case of metal blends, 
like the BL~2538 or BL~2740 indices. 

Recalling our previous physical arguments (see right panel of Fig.~\ref{fig:trend}), note that 
systematically, the loci of the maxima in the plots of Fig.~\ref{fig:idxteffmet} 
also depend on the metal content, this is, high metallicity indices peak at higher
temperatures. Because of this, in some indices, namely Fe~\textsc{ii}~2332, Fe~\textsc{ii}~2402,
and Mg~\textsc{ii}~2800, the effects of chemical composition are reversed after the
index saturates towards low $T_\mathrm{eff}$: at the lowest temperatures the
low metallicity indices are higher. The peak at different metallicities are
shifted by as much as 3000~K if we compare, for the index Fe~\textsc{ii}~2402, metal
rich indices with those of the lowest chemical composition. The variation of
the indices can be as high as a factor of three in the full metallicity
interval.

The three redder indices, Mg~\textsc{ii}~2852, Fe~\textsc{i}~3000, and,
notably, BL~3096  display marginal variations with chemical composition. This
property seems promising for separating the effects of age and metallicity in
integrated spectra.
 
\subsection{Effects of instrumental resolution}
\label{sec:resolution}
An important test that has to be conducted and quite often is neglected is
the analisys of sensitivity of spectral indices to instrumental
resolution. The comparison of theoretical vs. observed indices (either stellar
or integrated) or the analyses of indices measured from spectra from data sets
collected with different instrumentation should be carried out with data at
the same resolution. The risk of not doing so can be, for instance, the
incorrect segregation of potential useful indices. The changes due to spectral
resolution might have important implications in the calculation of
spectroscopic indices, in particular if the indices are used in the study of
the integrated properties of galaxies whose spectra are intrinsically broadened
by the motions of the stellar component.
 
\begin{figure}[t]
\begin{center}
\begin{tabular}{llll}
\resizebox{4cm}{!}{\includegraphics{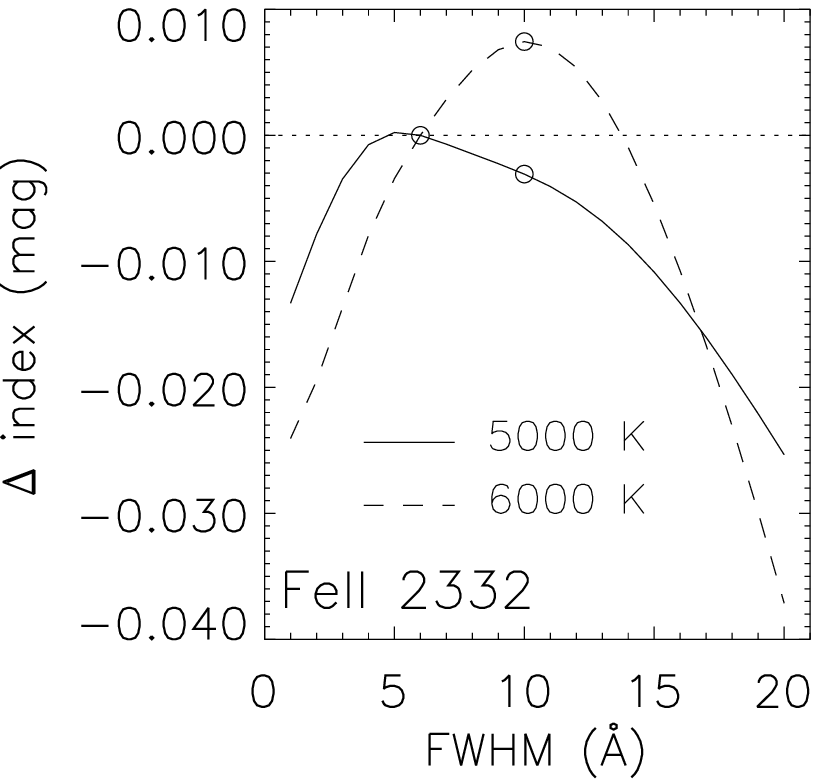}}   &
\resizebox{4cm}{!}{\includegraphics{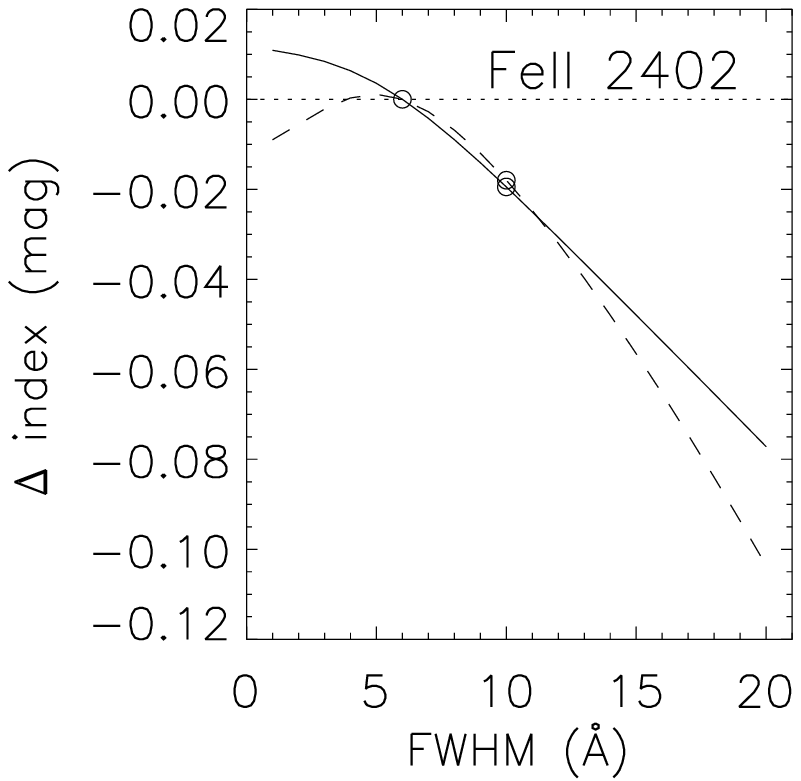}}   &
\resizebox{4cm}{!}{\includegraphics{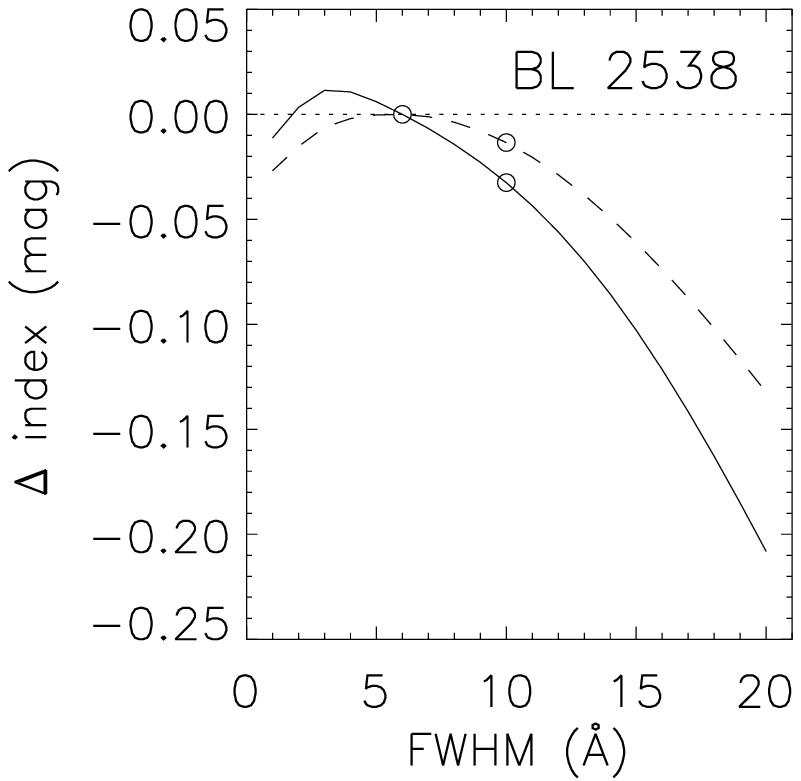}}     &
\resizebox{4cm}{!}{\includegraphics{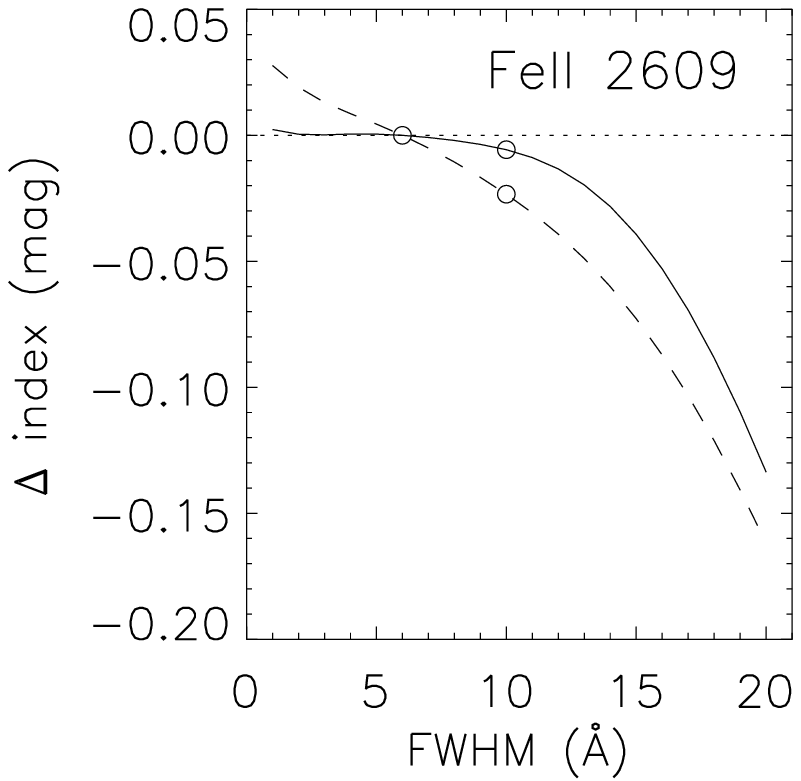}}   \\
\resizebox{4cm}{!}{\includegraphics{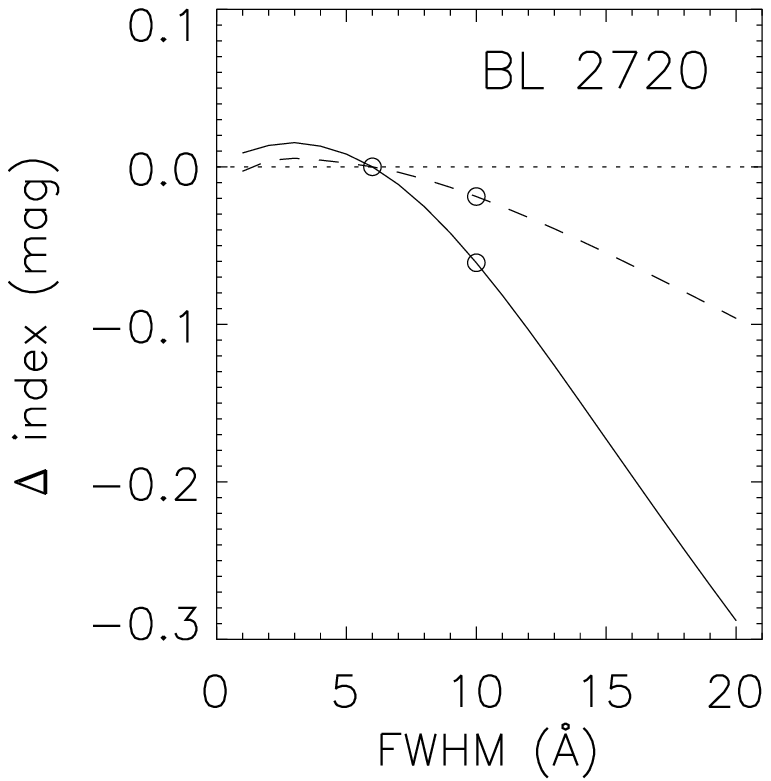}}   &
\resizebox{4cm}{!}{\includegraphics{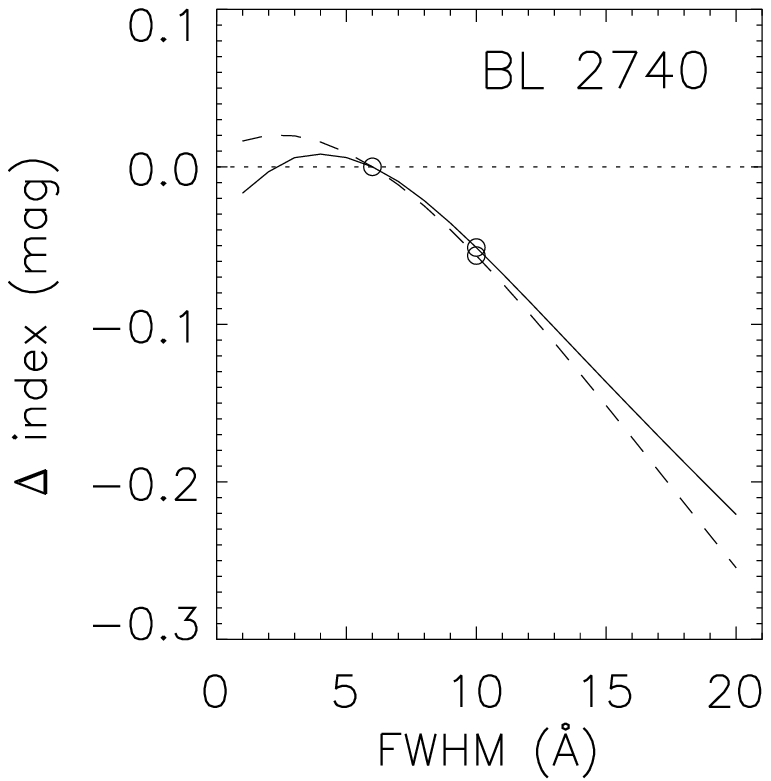}}     &
\resizebox{4cm}{!}{\includegraphics{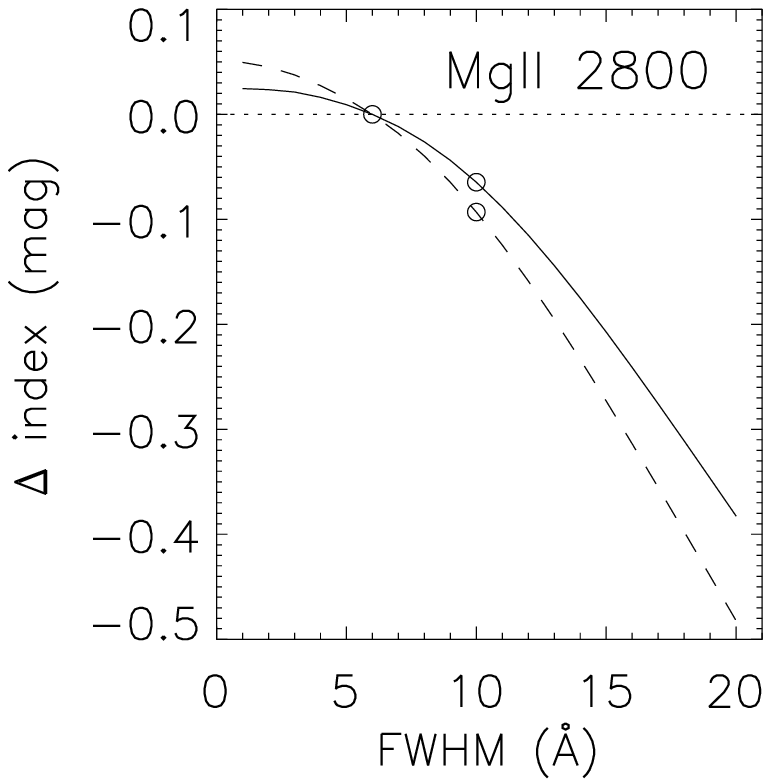}}   &
\resizebox{4cm}{!}{\includegraphics{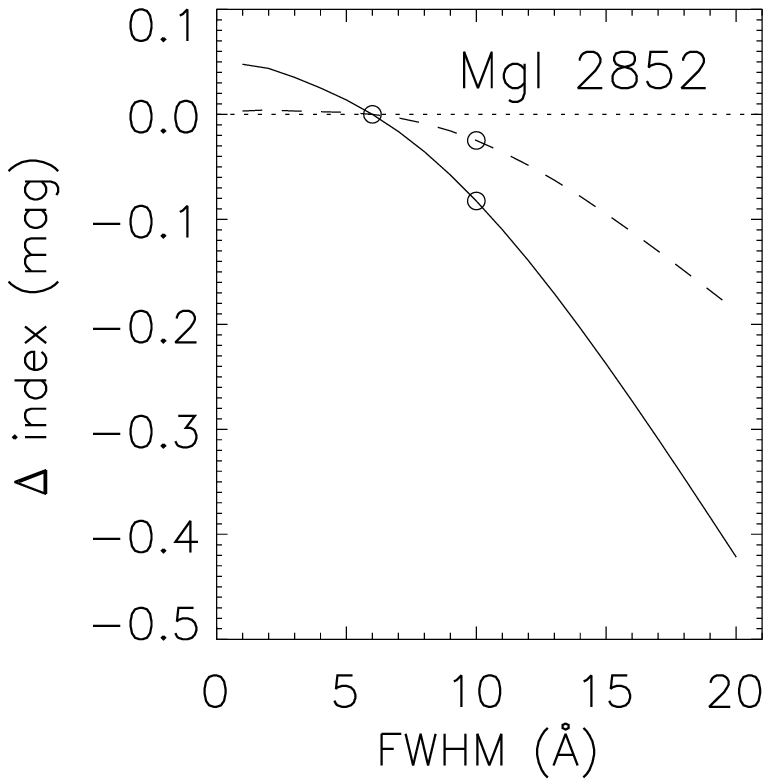}}    \\
\resizebox{4cm}{!}{\includegraphics{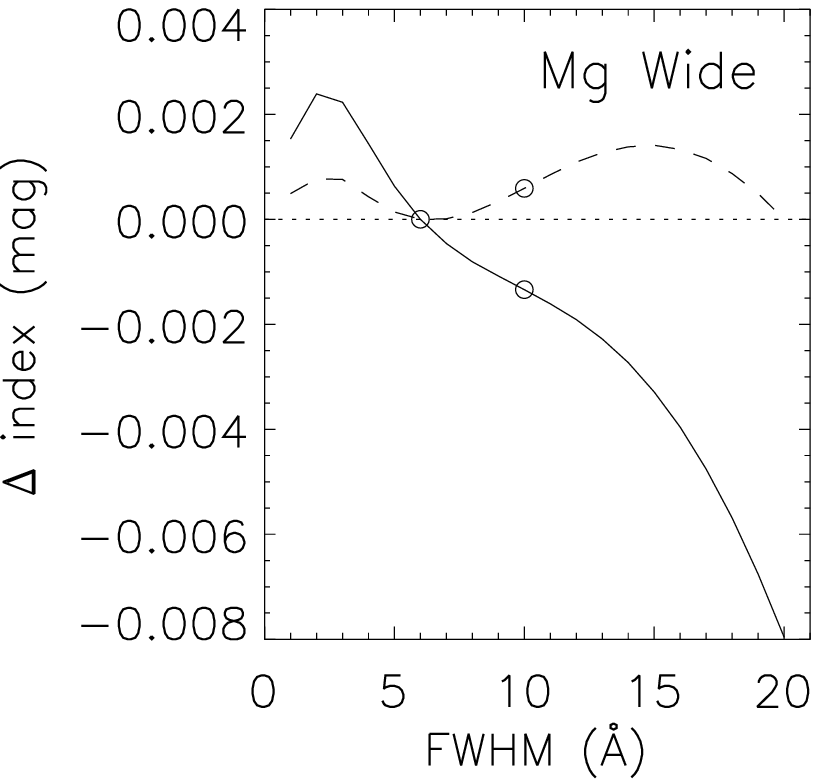}}     &
\resizebox{4cm}{!}{\includegraphics{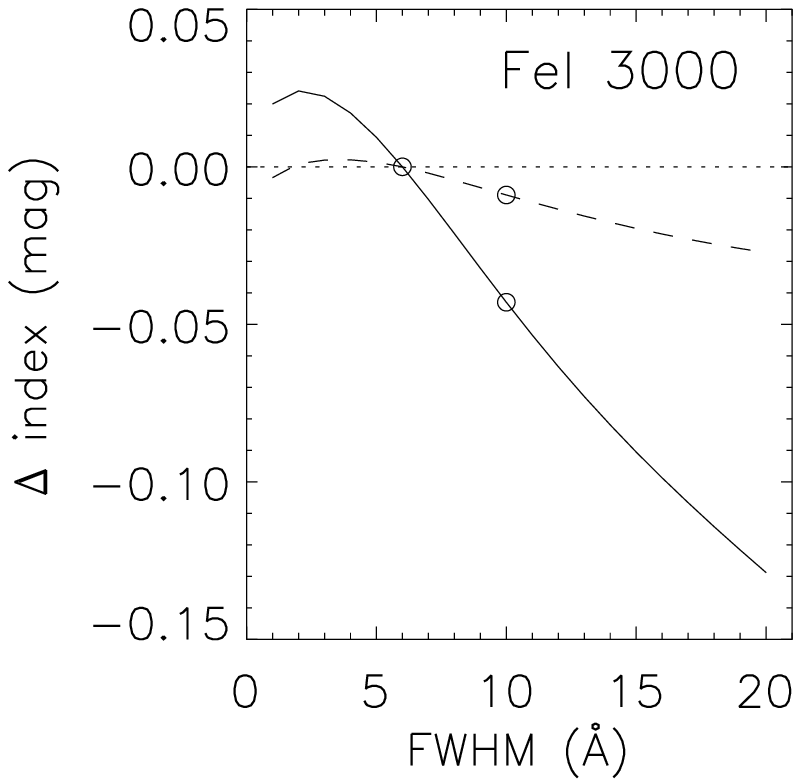}}    &
\resizebox{4cm}{!}{\includegraphics{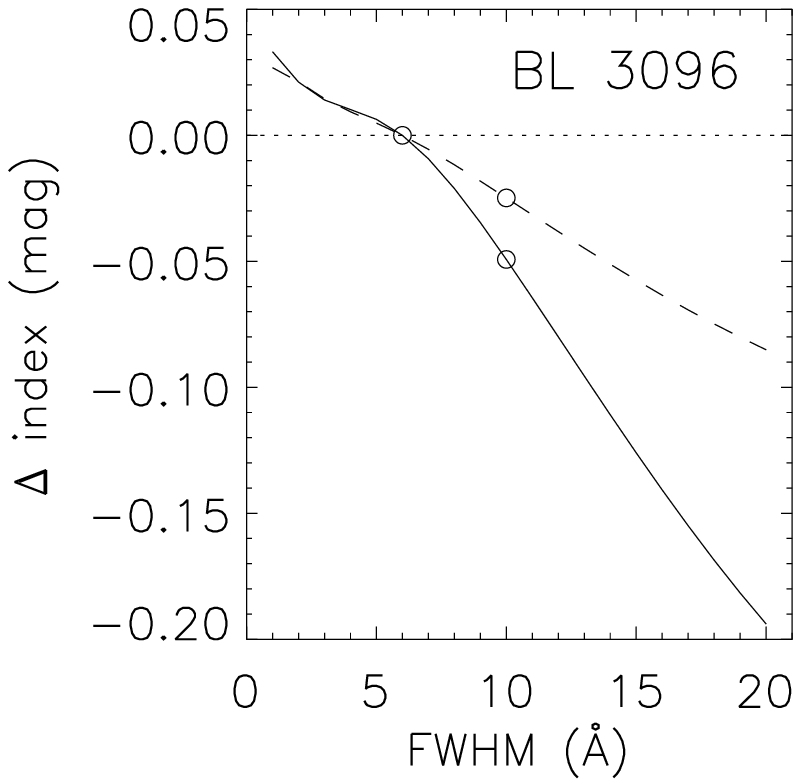}}     &
\resizebox{4cm}{!}{\includegraphics{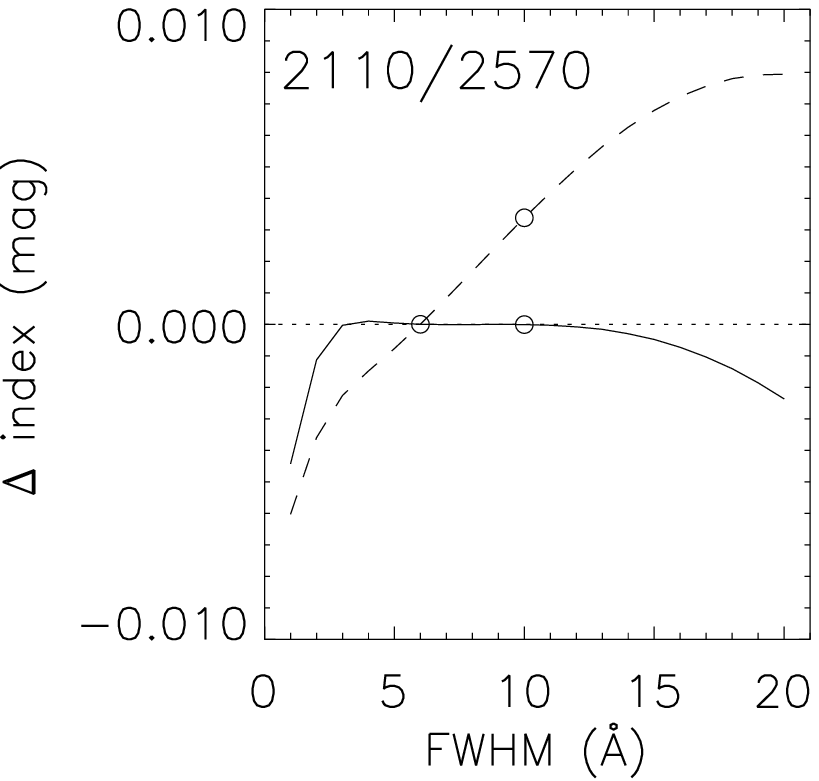}} \\
\resizebox{4cm}{!}{\includegraphics{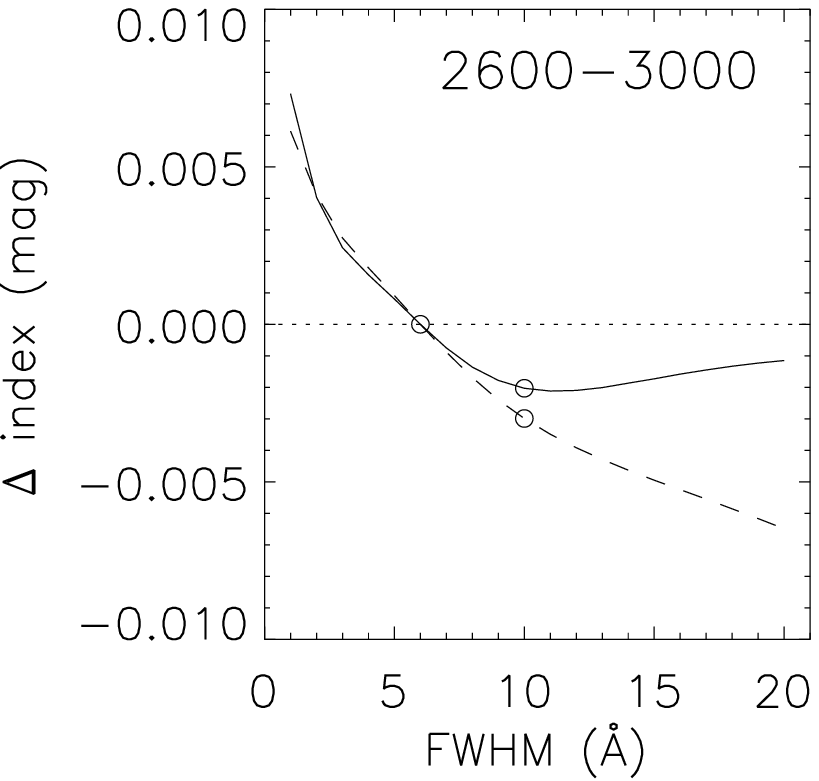}} &
\resizebox{4cm}{!}{\includegraphics{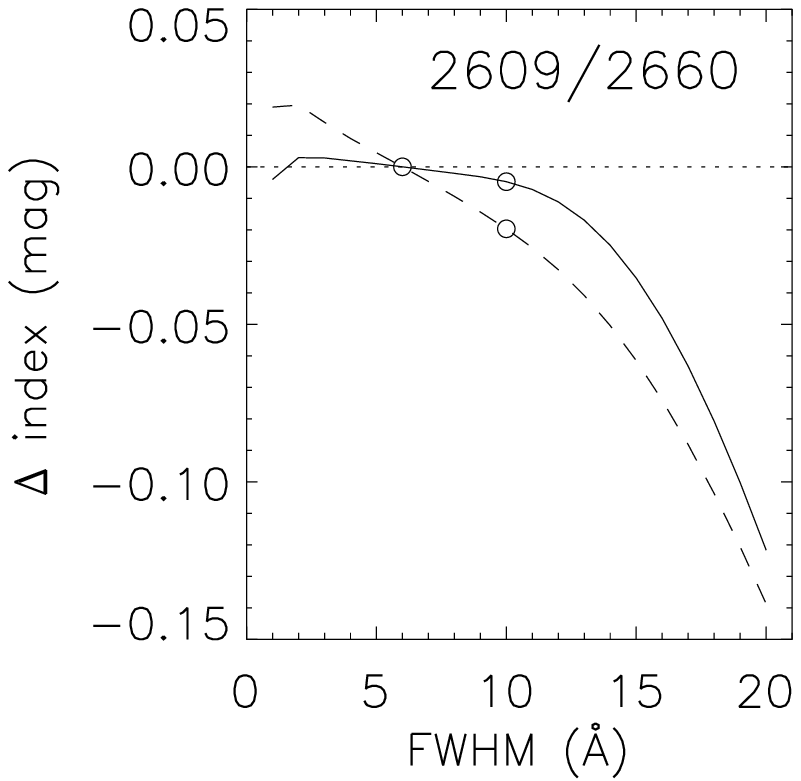}} &
\resizebox{4cm}{!}{\includegraphics{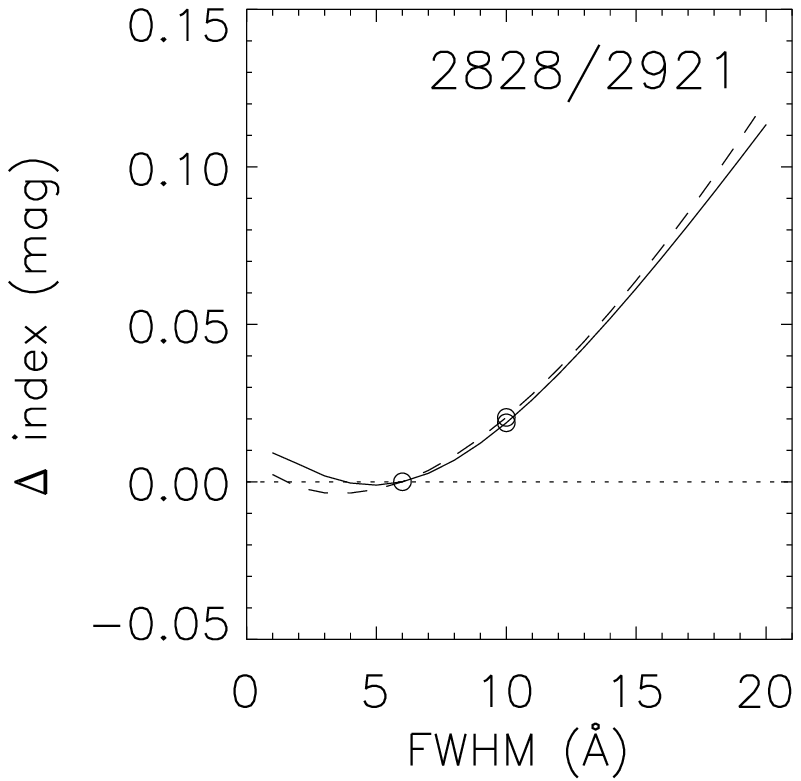}} &
\resizebox{4cm}{!}{\includegraphics{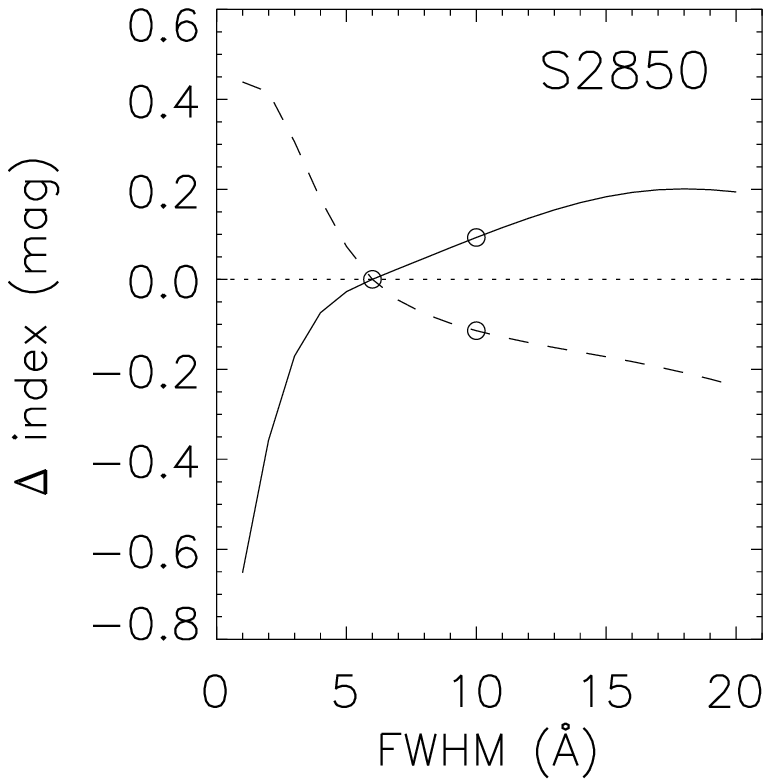}}      \\
\resizebox{4cm}{!}{\includegraphics{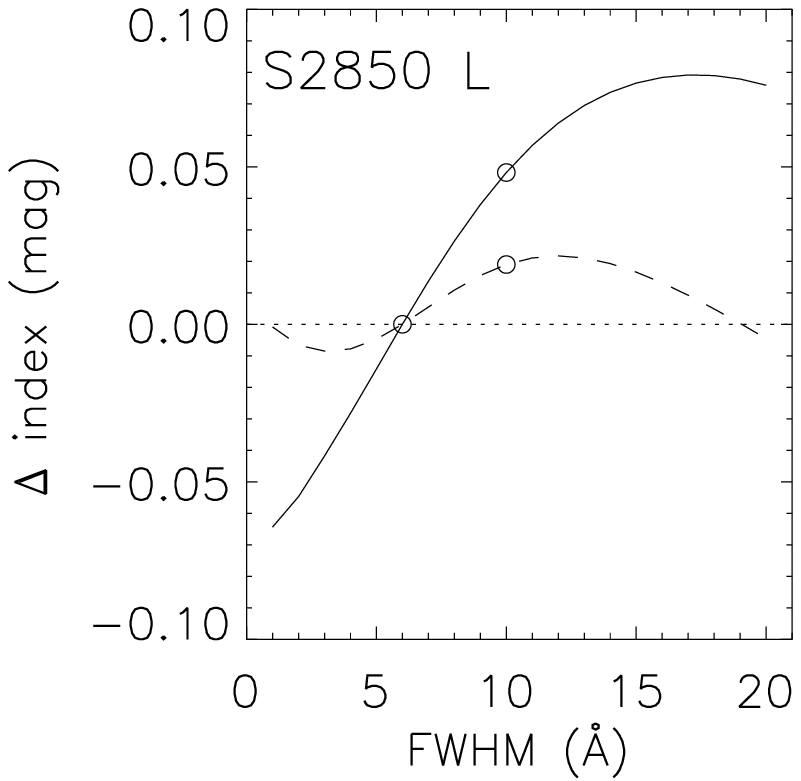}}    \\
\end{tabular}
\caption{Effects of instrumental resolution of index values. The y-axis
  indicates the index residual with respect to their value at the nominal IUE
  resolution (6~\AA) illustrated as a horizontal dotted line and an open
  circle at the 6~\AA--0~mag intercept. We have also denoted with an open
  circle index values at 10~\AA\ resolution compatible to the widely used
  Kurucz public fluxes. Solid and dashed lines correspond to indices for
  $T_\mathrm{eff}$=5000 and 6000~K synthetic spectra,
  respectively. \label{fig:idxres}}
\end{center}
\end{figure}

In Figure~\ref{fig:idxres} we show the effects of resolution on the line and
continuum indices discussed in the previous section. In the {\it y}-axis we
indicate the index difference with respect to its value
at a resolution of 6~\AA. For the analysis we have considered two synthetic
spectra of $\log{g} = 4.0$, [M/H]$=+0.0$ and $T_\mathrm{eff}= 5000$ and 6000~K
(represented by solid and dashed lines, respectively), temperatures at which
most indices reach their maxima. Data points correspond to indices measured in
spectra after applying Gaussian filters for differents FWHM ranging from 1 to
20~\AA\ with a step of 1.0~\AA. The open circles in each panel indicate the
position of the reference IUE resolution (6~\AA) and the resolution of the
Kurucz public library of fluxes \citep{Kurucz93}. In general, one would expect that resolution effects will be modulated by the
width of the bands, in the sense that broader bands will be less sensitive to
resolution. In a similar way, effects of resolution will depend on the
intrinsic width of the spectral lines, this is, whether a line is fully
embraced or not within the central bandpass. Let us first comment on the line
indices.

We can identify several properties of the effects of resolution on index
values. First, all line indices have their lowest value at the lowest
resolution (20~\AA). The only exception is Mg~Wide, which, however, displays a
marginal variation of less  0.002 mag between 6 and 10~\AA. Second, all
indices have their peak values either at the highest resolution or at about
2--4~\AA. This is most probably due to the role that features very near to the
limit of the bands (either inside and outside) play in modulating an
index. Third, with the exception of the index BL~3096, all indices vary up to
40\% in the full resolution range. BL~3096 doubles its value from 1
to 20~\AA, and changes as much as 0.05 mag (about 20\%) in the 6 to 10~\AA\ interval.

Concerning the continuum indices we find a wide diversity of behaviors,
nevertheless in four of them (2120/2570, 2600-3000, 2609/2660, and S2850L) the
effects of resolution are very weak within the full resolution range, 
while for the remaning two, 2828/2921 and S2850, values change
up to 0.12 mag and 0.8 mag (30 and 40\%), respectively. Note that this latter 
couple of indices are those defined with the narrowest bands. It is important to stress that the exercise shown here is based only on two
model spectra. At other temperatures and chemical compositions the effects
might decrease or be enhanced.
 

\section{Empirical vs Theoretical Indices}
\label{sec:emp_vs_th}
With the goal of quantitatively compare theoretical indices with indices 
derived from observed data, we have adopted the database described in Paper~I to
where the reader is referred for more details. In summary, our working sample
is based on the \citet{Wu83} atlas of IUE low resolution spectra and the
cool star extension by \citet{Fanelli90}. A sub sample of 111 stars was
secured by imposing that at least one full set of atmospheric parameters be
available in \citet{Cayrel97}.
  
Keeping in mind that there exist differences in the image extraction processes
used in the original \citet{Wu83} atlas and our IUE-INES data set, we
have, as a first step, verified the extent to which new indices measured in
re-processed IUE spectra are compatible with those reported by \citet{Fanelli90}. 
The overall result is that most indices display a one to one
correlation with the exception of Fe~\textsc{ii}~2402 where there is a quite high
dispersion for points corresponding to giant stars. The most plausible
explanations for these  differences are, on one side, the different absolute
flux calibration curves used in {\it INES} \citep{Gonzalezriestra01} and {\it IUESIPS}
 \citep{Bohlinetal80} and, on the
other side, the fact that we used all available images of good quality (in
some cases more than 50) while \citet{Wu83} considered only a few. In
addition to this, one has to consider that spectra of cool stars are often
quite noisy since their intrinsic UV flux descreases drastically and that at
the short wavelength edge  the flux is close to the LWP/R cameras sensitivity
limits. With the indices computed in the re-processed IUE images we have
defined the standard system.

\begin{figure}[t]
\begin{center}
\begin{tabular}{cc}
\resizebox{7cm}{!}{\includegraphics{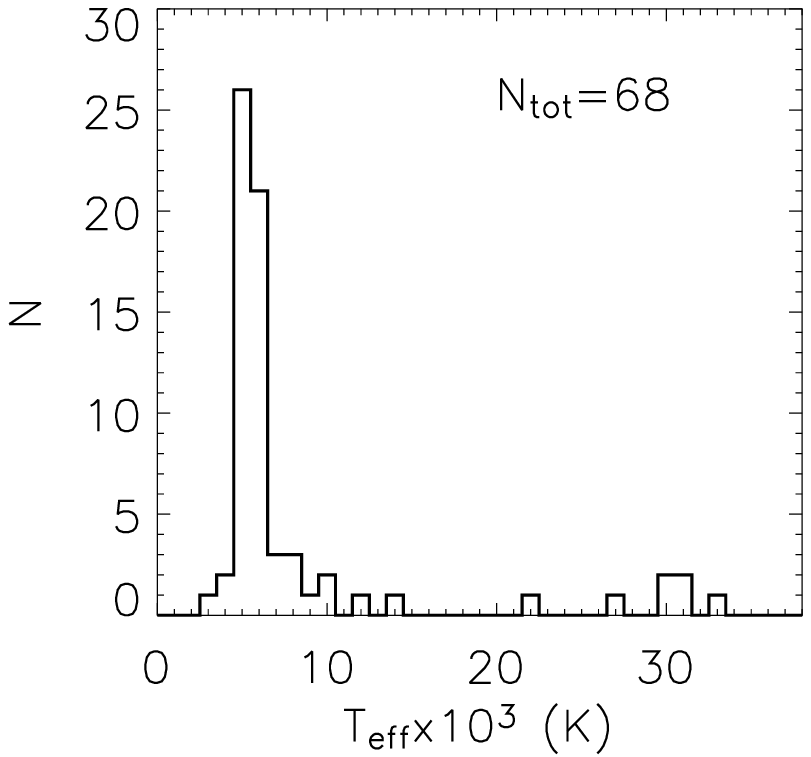}}   &
\resizebox{7cm}{!}{\includegraphics{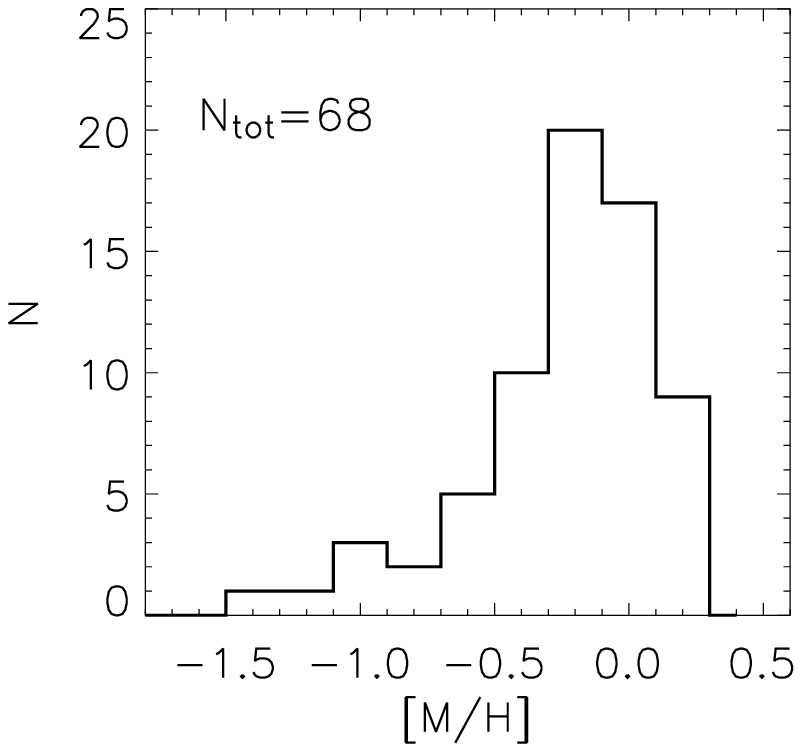}}   \\
\end{tabular}
\caption{The distributions in $T_\mathrm{eff}$ and [M/H] of the
    high-gravity IUE stellar sample that is used for the comparison of synthetic
    indices. The sample is mainly composed of G and F-type stars whose
    metallicity distribution (right panel) peaks at about the solar
    value. \label{fig:histo}}
\end{center}
\end{figure}

From the theoretical side, we have created a synthetic data set through a
three-linear interpolation of the flux, divided by the corresponding 
$\sigma T_\mathrm{eff}^4$, of
the UVBLUE spectra in the space ($T_\mathrm{eff}$, $\log{g}$, [M/H]). For each
star we adopt the parameters listed in Table~2 of Paper~I.   As in the case of
the synthetic SEDs presented in the previous section, these spectra were degraded 
to match IUE resolution (6~\AA). It is important to mention that we have imposed one restriction for the
comparison: we have excluded low gravity objects ($\log{g} \leq 3.5$) since,
on one side, giant and supergiant stars show the largest discrepancies between
observed and theoretical SEDs according to Paper~I and, on the other side,
intrinsic UV fluxes of low gravity objects at the short wavelength edge of IUE 
are lower that the high gravity
counterparts. Additionally, evolved objects of low surface gravity are not
expected to significantly contribute to the mid-UV flux of early-type systems,
one of the main foreseeable applications of synthetic UV indices. The final
sample for our comparison is composed of 68 objects whose distribution in
$T_\mathrm{eff}$ and [M/H] is displayed in Figure~\ref{fig:histo}. The restriction on wavelength for the material presented in this paper
($\lambda \geq 2300$~\AA, except for the spectral break index 2110/2570)
ensures a higher signal to noise in observed data. Note also that no
restriction on chemical composition was imposed.

\begin{figure}[t]
\begin{center}
\begin{tabular}{llll}
\resizebox{4cm}{!}{\includegraphics{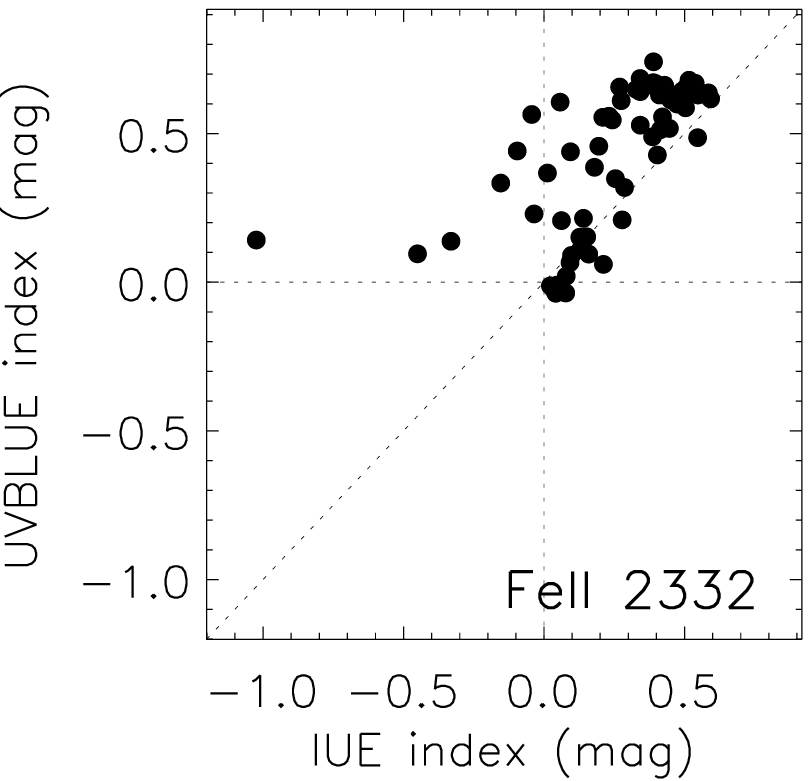}}   &
\resizebox{4cm}{!}{\includegraphics{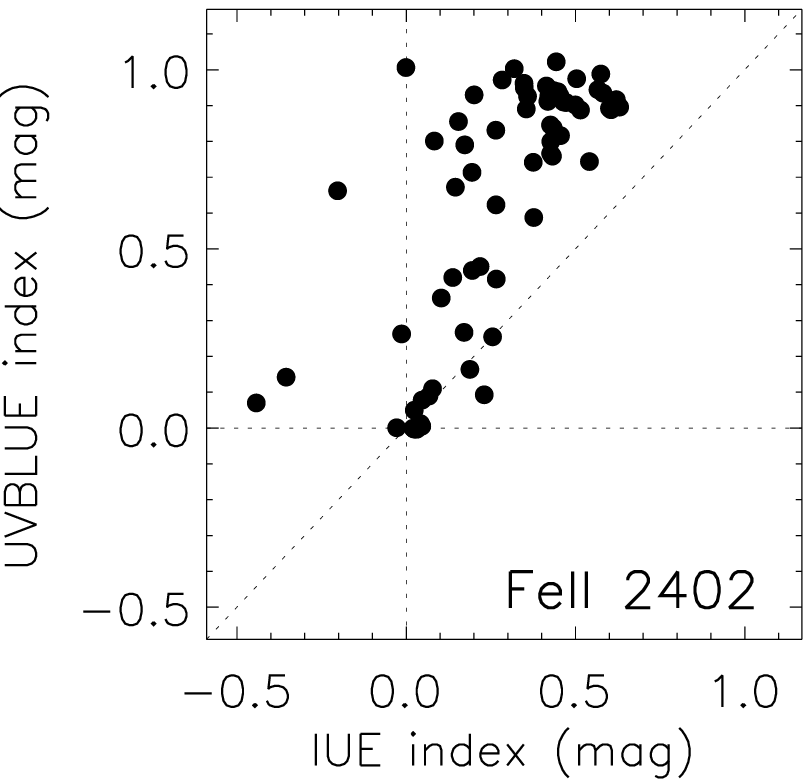}}   &
\resizebox{4cm}{!}{\includegraphics{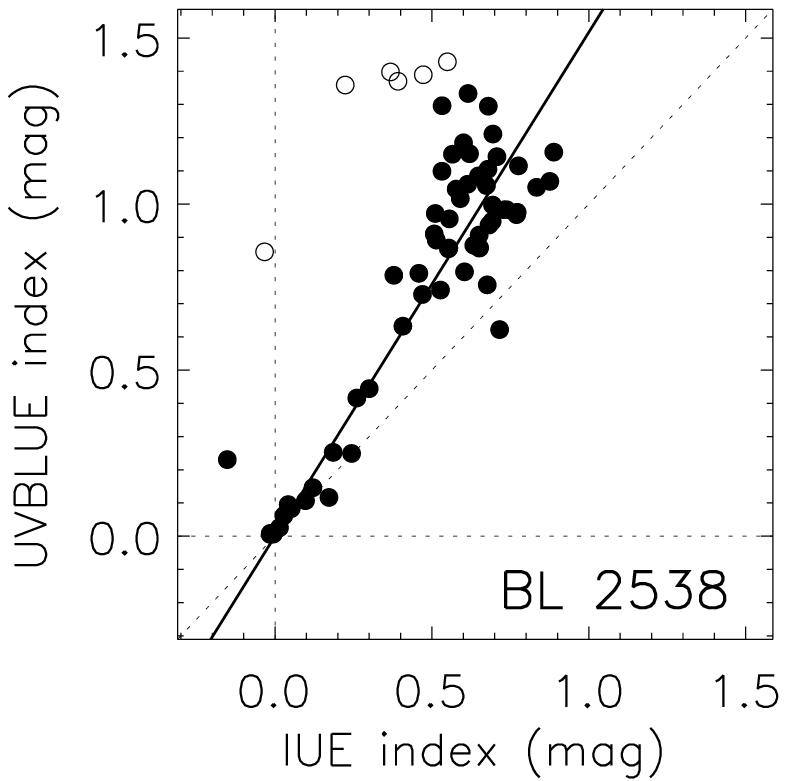}}     &
\resizebox{4cm}{!}{\includegraphics{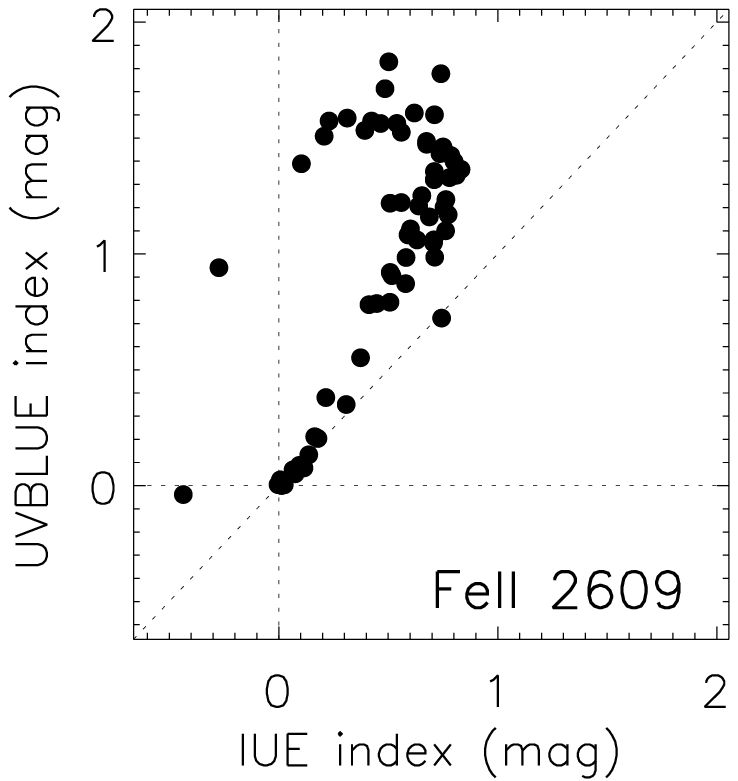}}   \\
\resizebox{4cm}{!}{\includegraphics{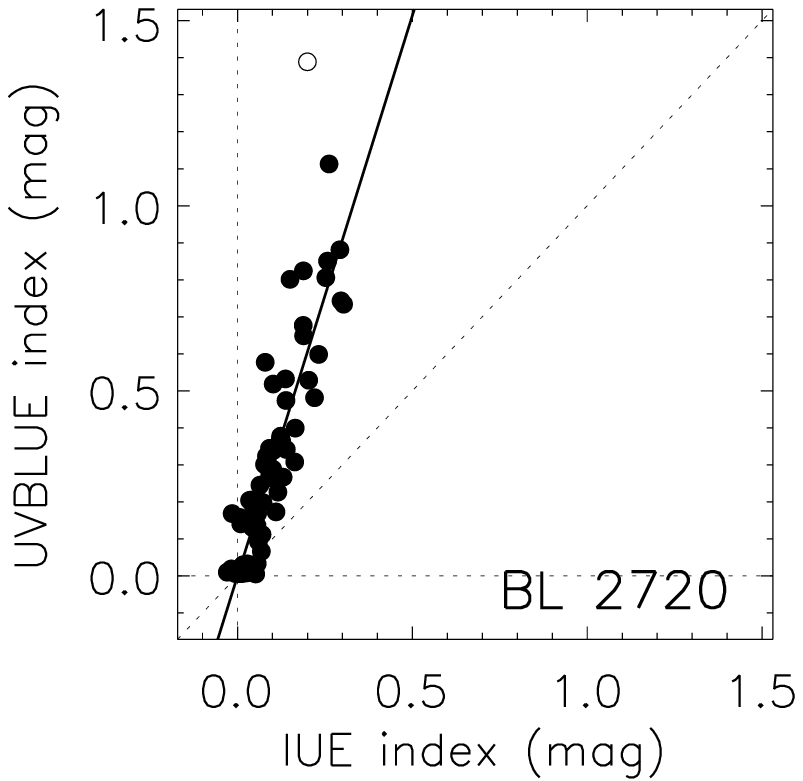}}   &
\resizebox{4cm}{!}{\includegraphics{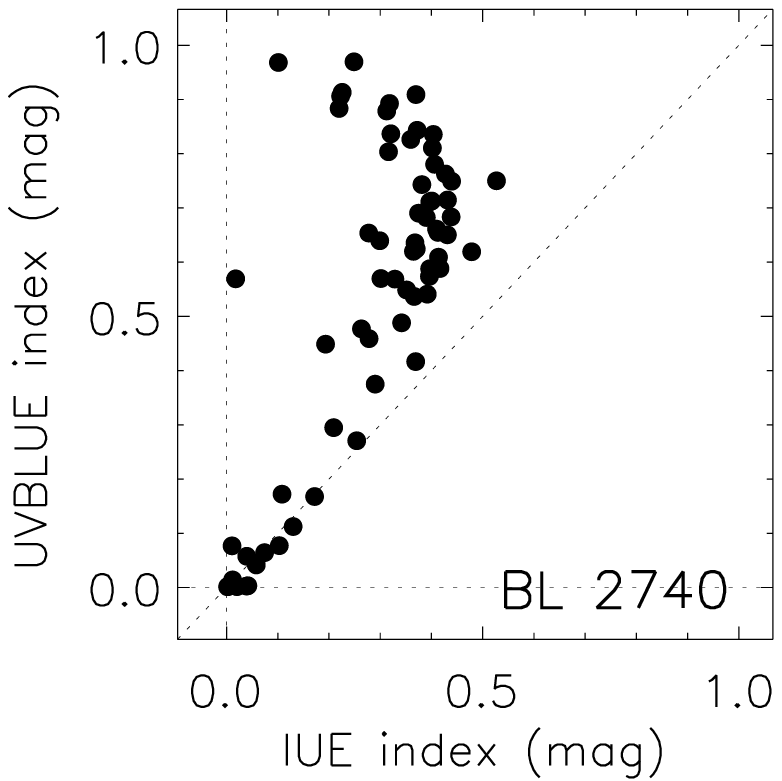}}     &
\resizebox{4cm}{!}{\includegraphics{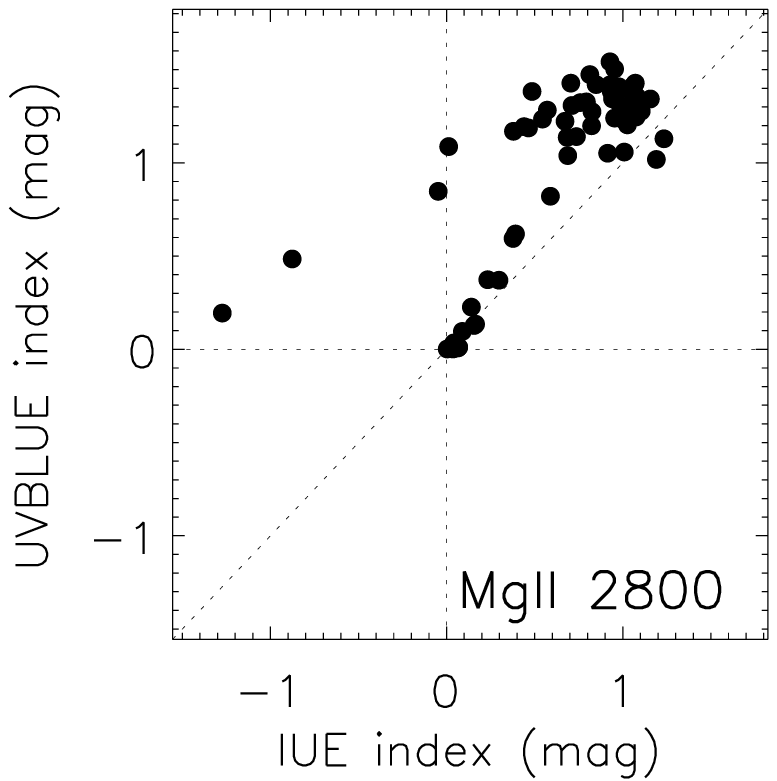}}   &
\resizebox{4cm}{!}{\includegraphics{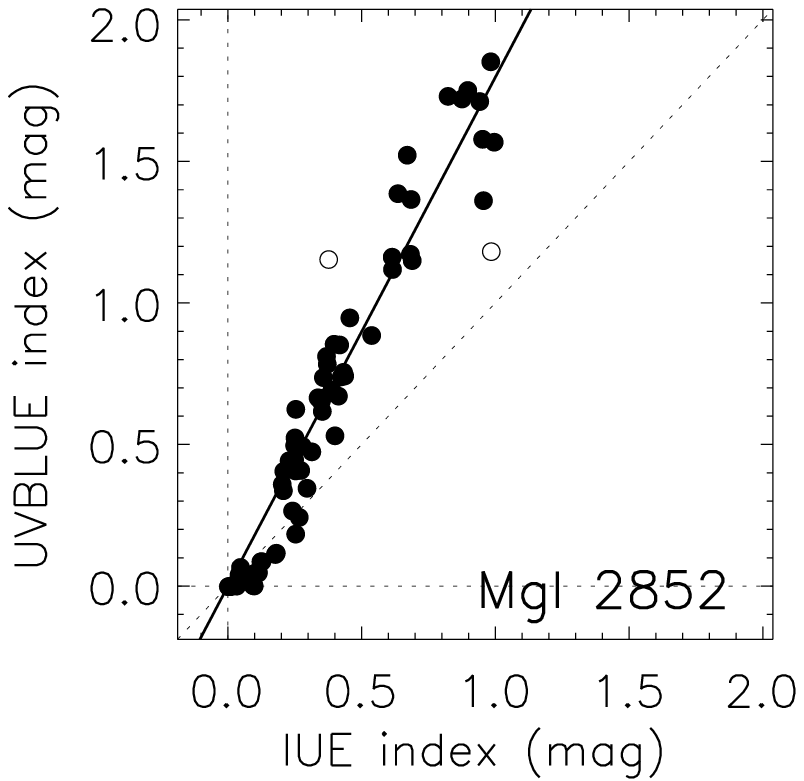}}    \\
\resizebox{4cm}{!}{\includegraphics{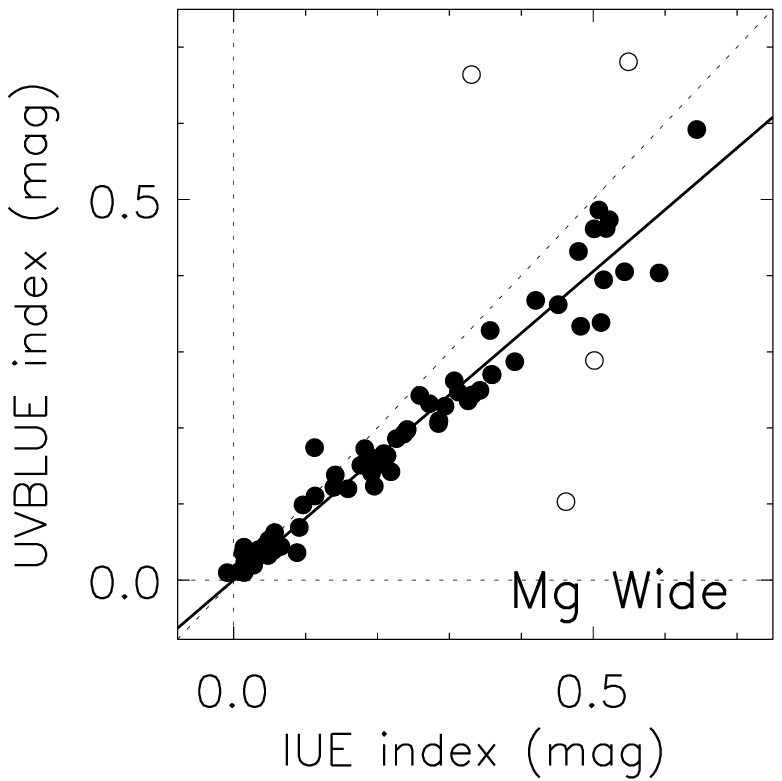}}     &
\resizebox{4cm}{!}{\includegraphics{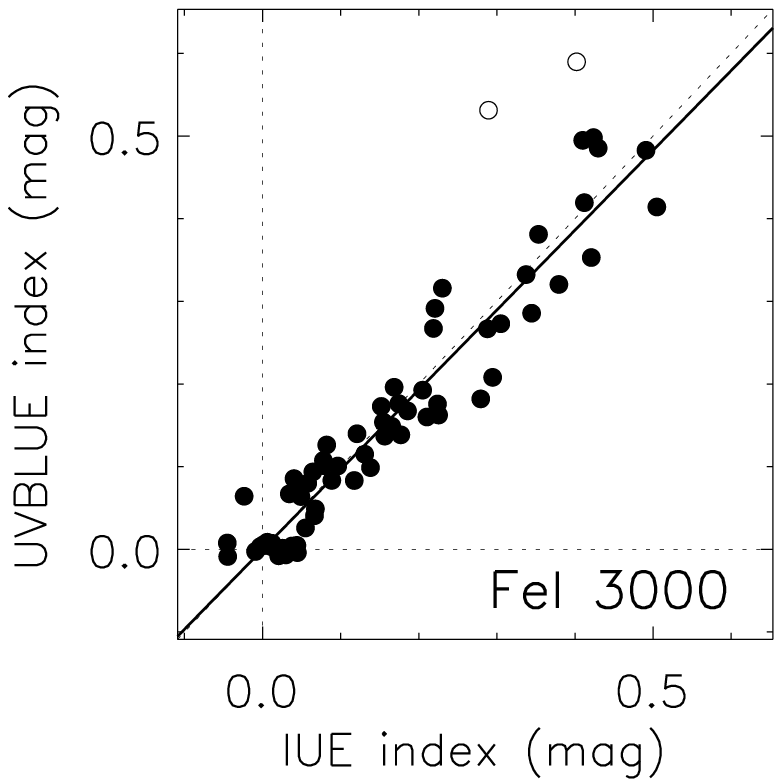}}    &
\resizebox{4cm}{!}{\includegraphics{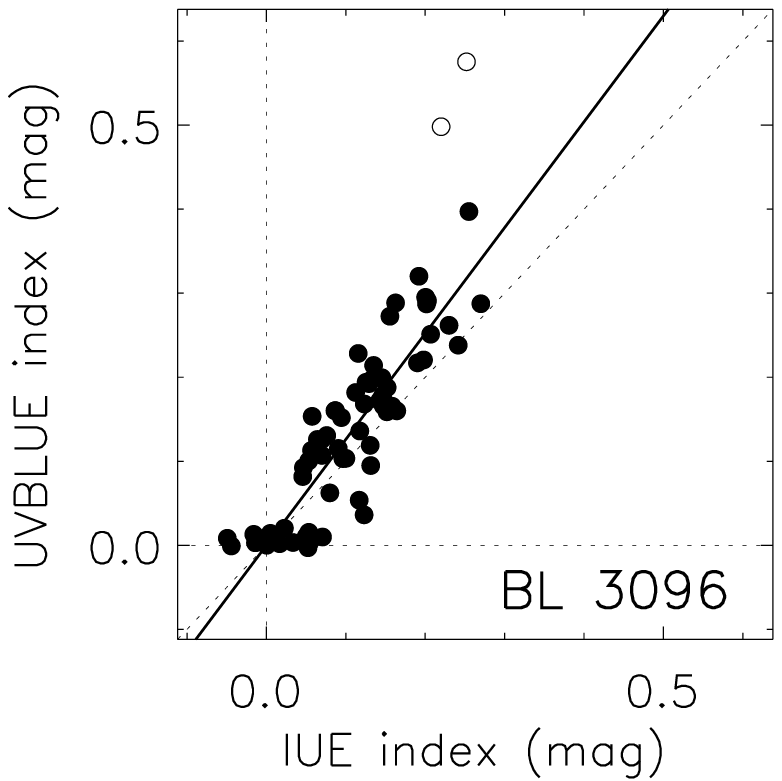}}     &
\resizebox{4cm}{!}{\includegraphics{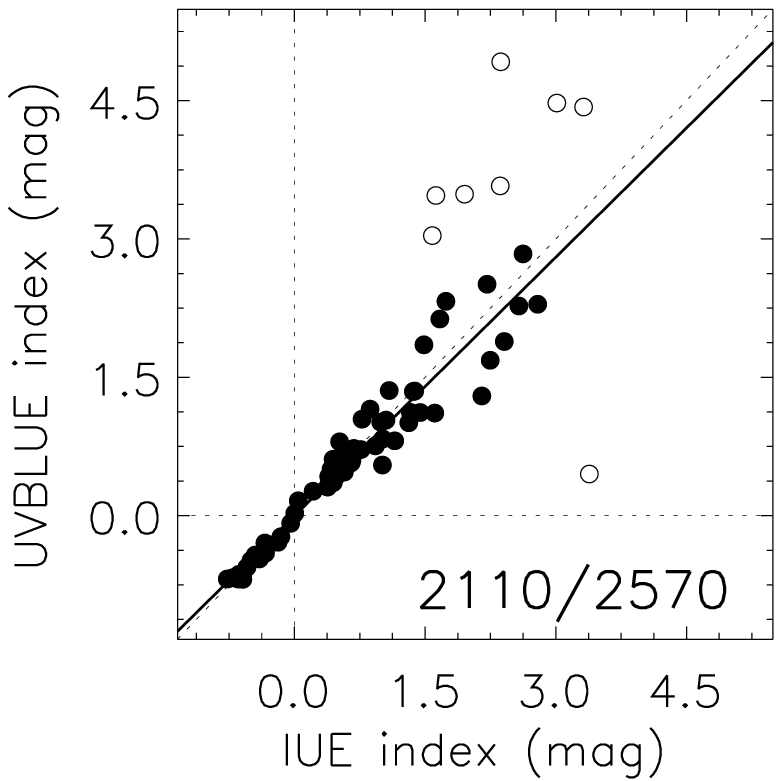}} \\
\resizebox{4cm}{!}{\includegraphics{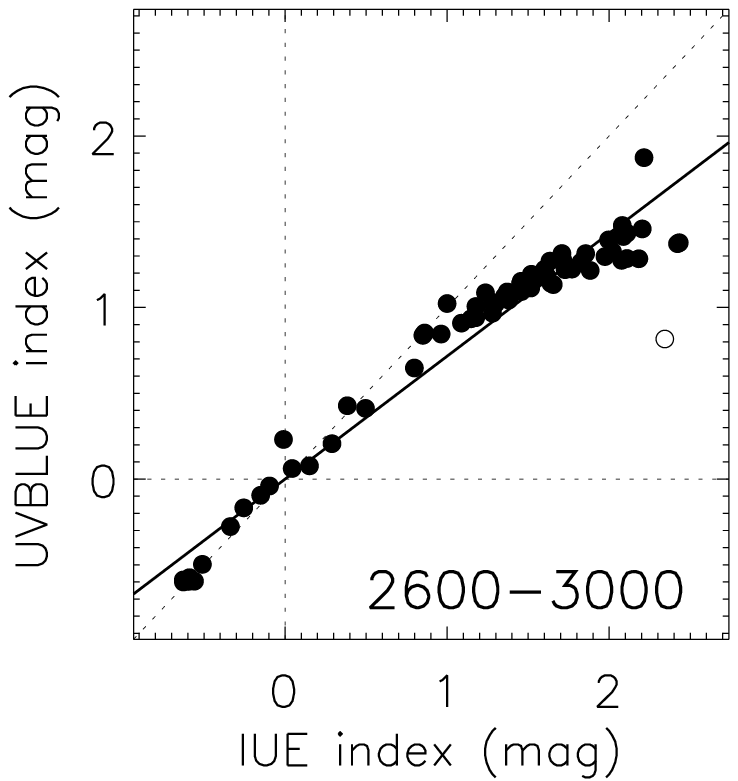}} &
\resizebox{4cm}{!}{\includegraphics{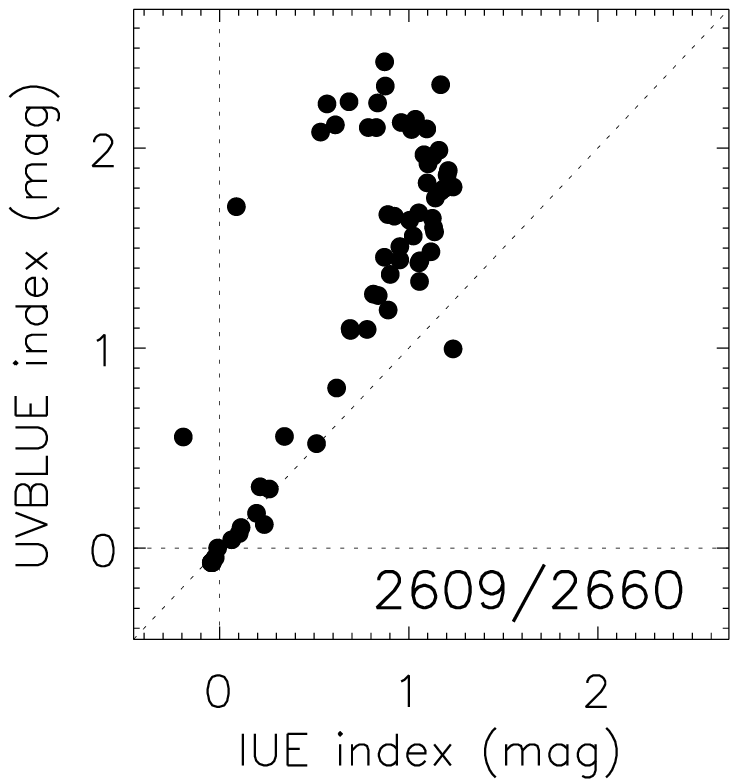}} &
\resizebox{4cm}{!}{\includegraphics{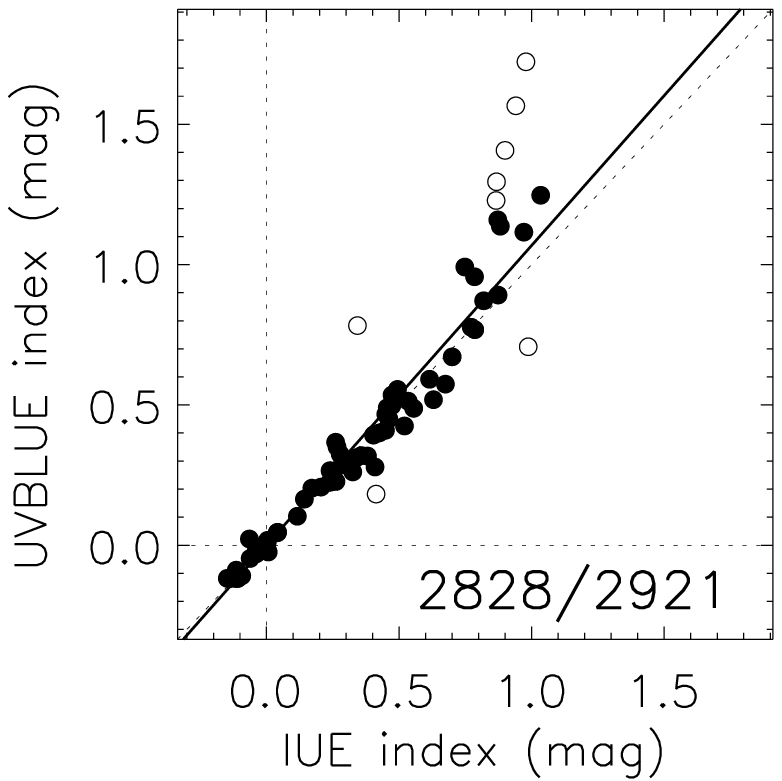}} &
\resizebox{4cm}{!}{\includegraphics{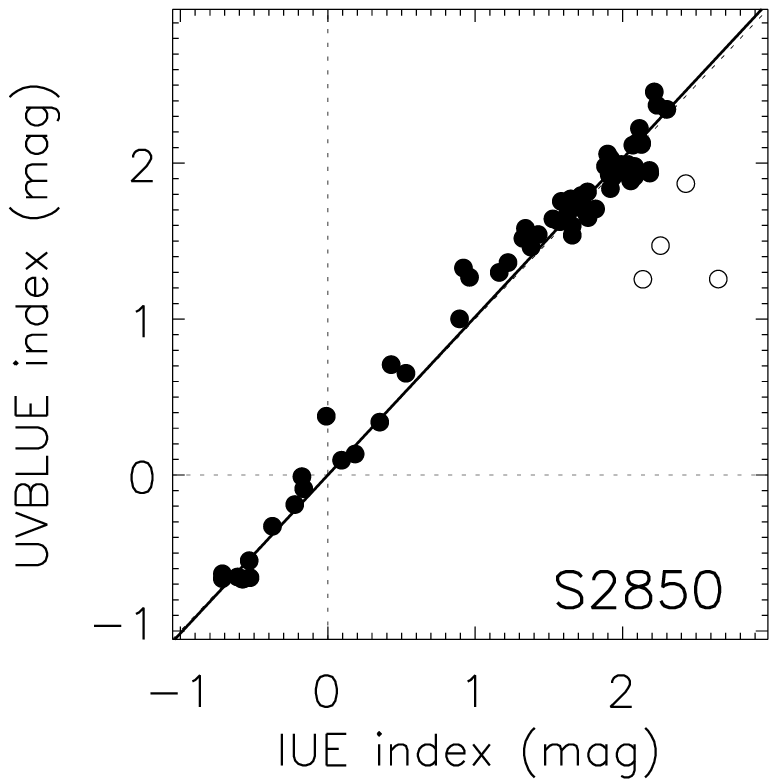}}      \\
\resizebox{4cm}{!}{\includegraphics{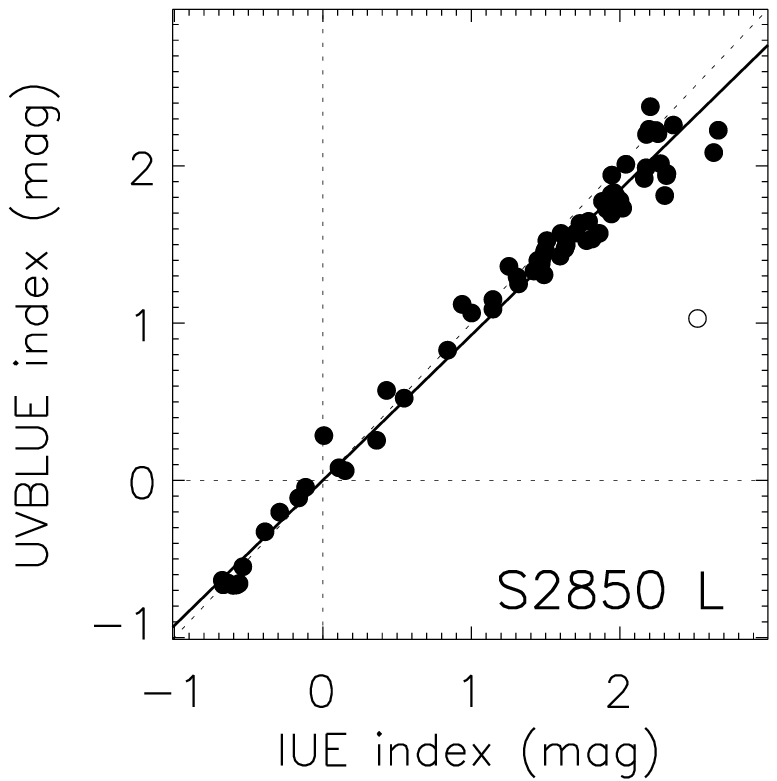}}    \\
\end{tabular}
\caption{Synthetic vs. empirical indices for IUE high-gravity ($\log{g} >
  3.5$~dex) stars. Panels in this figure are constructed by measuring indices
  with the definitions given in Table~\ref{tab:indices} and linearly
  interpolated UVBLUE spectra for the parameters listed in Paper~I. The
  straight dotted line illustrates the location of the one-to-one correlation
  (slope unity). The solid lines in eleven panels  depict the least square
  linear fits for these indices. Empty circles denote the rejected data points
  as explained in the text (see Sec.~\ref{sec:calib} for details)
  \label{fig:syniue}}
\end{center}
\end{figure}

The results of the comparison are displayed in Figure~\ref{fig:syniue}, where
we plot the synthetic indices ($y$-axis) and IUE-indices ($x$-axis) together
with a reference 45$^\circ$ dotted line. Ten indices are overestimated by the
results of classical model atmospheres. Of these indices two of them 
(Fe~\textsc{ii}~2332 and Fe~\textsc{ii}~2402) represent the worst cases with a
large scatter over the whole index range. All the remaining overestimated
indices (BL~2538, Fe~\textsc{ii}~2609, BL~2720, BL~2740, Mg~\textsc{ii}~2800,
Mg~\textsc{i}~2852, BL~3096, and the spectral break 2609/2660) exhibit a clear
linear correlation; however, in five of them there is a notable enhancement of
scattering on the points for low temperatures stars, not so for the indices
BL~2720, Mg~\textsc{i}~2852, and BL~3096. Additionally, it is interesting to
note that in three indices with large scatter at low $T_\mathrm{eff}$ the
linear relation appears to be truncated and the empirical indices show a
reversal, i.e. indices saturate and decrease after reaching their maxima
towards low temperatures. This situation is not modelled by the synthetic
indices. Two indices (Mg~Wide and the continuum index 2600-3000) appear
slightly underestimated by theory. Finally, five indices appear properly
reproduced by UVBLUE indices: 2110/2570, Fe~\textsc{i}~3000, the spectral
break 2828/2921, and the continuum indices S2850 and S2850L.

Let us compare our results with those of Lotz, Ferguson \& Bohlin (2000, herefater LFB00) which represent, to our
knowledge, the only attempt so far to compare Kurucz-derived synthetic mid-UV
indices with observations. They present eight indices all of which are
included in our analysis, although, our definition of S2850 is different from that of
LFB00. They concluded that for five of these indices, namely,
Fe~\textsc{ii}~2402, Fe~\textsc{ii}~2609,  Mg~\textsc{ii}~2800,
Mg~\textsc{i}~2852, and Mg~Wide, values are overestimated by Kurucz
model fluxes. They found a good match for BL~2538 and for S2850 (the latter
for stellar models with $T_\mathrm{eff} \geq 4500$~K), while Fe~\textsc{i}~3000 is
underestimated by 0.15 mag. While our results agree for the first four
indices, for the rest we obtain contradicting behaviors. The BL~2538 index
appears to be also overestimated, particularly at low temperatures, Mg Wide is
somewhat underestimated, and our synthetic Fe~\textsc{i}~3000 and the slope
between 2600 and 3000~\AA\ match very well empirical indices. There are three
plausible explanations for the conflicting results. First, the fact that LFB00
measure theoretical indices from the set of Kurucz model fluxes compiled by
\citet*{Lejeune97} which are at a nominal resolution of 10~\AA\, while our
high resolution theoretical data set has been degraded to 6~\AA. As we have
explained in Section~\ref{sec:resolution}, this change in resolution can
account, at least partially, for the discrepancies. For instance, BL~2538 and
Fe~\textsc{i}~3000 decrease by about 0.04~mag after changing the resolution
from 6 to 10~\AA. Second, Kurucz model fluxes are calculated using the opacity
distribution functions (ODFs) while the UVBLUE SEDs account for detailed line
opacity. It is clear that the statistical nature of the ODFs dilutes the
effects of incomplete and/or incorrect line opacities, however, at the price
of preventing the possibility of producing spectra at high enough resolution 
\citep[see also the discussion in][]{Dorman03}. Third, in our comparison we
have accounted for the individual stellar parameters while LFB00 provide a
comparison of empirical indices obtained from mean stellar IUE spectra with
those measured from a single-gravity single-metallicity (solar) theoretical
flux. Even though the definition of the S2850 index in LFB00 and our work
differs, our comparison indicates that the slopes we measure in
UVBLUE-interpolated spectra closely reproduce empirical results.

Various agents more can be brandished to explain the discrepancies seen in 
Figure~\ref{fig:syniue}. As briefly mentioned by LFB00 and
listed in Paper~I as caveats of UVBLUE, these are: incomplete line lists,
which directly affect the heavily blanketed ultraviolet fluxes; departures
from LTE and chromospheric heating.  Additionally, and not explicitly mentioned in the
above investigations, is the fact that, even if we had a complete line list,
the line parameters (mainly the oscillator strengths, Van der Waals damping
constants and wavelengths) might be wrong. We believe that discrepancies in
many of the indices can be significantly reduced with the improvement of 
these three parameters, since, as demonstrated in Paper~I, individual lines
are systematically stronger in theoretical spectra.

\subsection{Comments on discrepant indices}

A thorough spectral line study and their effects on individual index bands
is necessary to unambiguously identify the agent(s) that provoke the
discrepancies in the comparison between observed and theoretical indices. Such
an analysis is beyond the scope of this work, however, we can speculate on the
most likely explanations.

{\it Mg~Wide and 2600-3000}: the underestimation of these two indices is
attributed to a lower opacity in their shared blue band which
partially includes the spectral region 2640--2700~\AA. In Paper~I we
have already pointed out that this region is poorly reproduced by model spectra; the
most plausible reason for this is the lack of many absorption lines in
the UVBLUE line list \citep{Peterson01}. Furthermore, in UVBLUE we did not
considered the so-called ``predicted" lines \citep{Kurucz92} which
affect the blanketing in this region. The reason for avoiding the use of lines
with non-laboratory atomic data is that UVBLUE is also intended for high
resolution analyses and the inclusion of unreliable data greatly increases
confusion for line identification. Moreover, for Mg~Wide these effects are
mixed with the overvalued strength of the Mg features comprised in its central
band. We do not expect any important contribution to the inconsistencies from
the red band since, on one side, we know that within the mid-UV spectral
interval line blanketing effects are enhanced towards shorter wavelengths. On
the other side, the red and middle bands and partially the blue band of the
index Fe~\textsc{i}~3000 are embraced by the red band of Mg~Wide and 2600-3000, and
this index is well reproduced.

{\it Fe~\textsc{ii}~2609, BL~2720, BL~2740}: these indices, which share one
side band, are also affected by the incompatibilities between the theoretical
and observed SEDs in the region 2640--2720~\AA\ which result in the largest
deviations from the one-to-one reference line.

{\it Fe~\textsc{ii}~2332, Fe~\textsc{ii}~2402, Mg~\textsc{i}~2852}: these
indices are defined in regions where part of the discrepancies can be ascribed
to differences in the individual feature intensities themselves. An additional
aspect that should be taken into account is that the discrepant results of the
two bluer indices might also be ascribed to the low quality of observational
data in the wavelength interval used to define them. This is supported by the
large scatter present even at low index values. Several authors have
concentrated on redder indices because of this potential limitation.

{\it Mg~\textsc{ii}~2800}: the prominent Mg~\textsc{ii} line at about 2800~\AA\ is filled by
chromospheric emission. The different extents to which stars have
chromospheric activity could provide the observed dispersion in the points,
particularly at low $T_\mathrm{eff}$ (i.e. at the largest value of the
index). Even though we knew in advance that this feature could not be modelled
because of the non-thermal heating of the upper atmosphere, we decided to
include it for the sake of completness and to explore its behavior,
particularly in terms of chemical abundance. In fact we found that different
metallicities peak at different temperatures and therefore it could partially
explain its paradoxical behavior in stellar populations where the index is
smaller with increasing metallicity. If this is true one should explore in
more detail the less studied indices Fe~\textsc{ii}~2332 and
Fe~\textsc{ii}~2402 where these effects arise more pronounced.
 
{\it BL~3096}: this index is marginally overestimated by UVBLUE model fluxes
until it reaches a value of about 0.25 which roughly corresponds to
intermediate K-type stars. Larger indices (i.e. cooler stars) are apparently
not reproduced; however, there are too few points to draw any conclusion.

It is important to stress that to reach a better agreement between
  theoretical and the empirical results, an improvement of the line
  parameters is much needed. We anticipate that we have initiated a thorough 
  test of the line parameters
for the strongest lines in the mid-UV range (excluding those that are severely
affected by chromospheric emission).


\section{Re-calibration of discrepant theoretical indices}
\label{sec:calib}

We have seen that five out of seventeen empirical indices are adequately
reproduced by their synthetic counterparts. This result invites to a thorough
analysis of the main agents affecting the UV spectra in late-type
stars. However, we can still extract useful information from the indices that
display a clear linear correlation between theory and observations. 
This can be done by calibrating the synthetic indices to
match empirical ones in a similar way \citet*{Chavez96} did
for optically defined indices. We recall that no restriction has been imposed
on metallicity. Hence, the resulting calibration should be in principle
applicable to the full chemical composition and effective temperature ranges
involved in our IUE database, thanks to the improved homogeneity in the
parameter space provided by the theoretical spectra.

The general underlying goal of the transformation of a set of data to a
standard system could be manyfold. For example, it is required when two data
sets have not been obtained with the same instrument, i.e. have different
resolution. When the flux calibration in both sets is different or has not
been conducted at all in one or both and therefore it is necessary to match
the different detector responses.
A standard system is usually conformed by a sufficiently large sample of
stellar observed spectra. This transformation has been a common practice in
the analysis of, for example, the collection of optical indices defined by the
Lick group. For the case of mid-UV indices we have defined, as
mentioned previously, the mid-UV standard system as the set of indices
measured from the re-processed IUE data.

In our particular case, the transformation is aimed at transporting synthetic
indices to the expected empirical values. In Figure~\ref{fig:syniue} we have represented
with a solid line the least squares fit of the function
$I_\mathrm{syn} = b\,I_\mathrm{IUE}$ for 11 indices. We constrained
the fitted line to include the (0,0) point, i.e. both the theoretical
and empirical indices vanish at the same temperature. However, one can see
that, in any case, without this restriction the $y$-intercept would be very
close to zero.
The fitting process has been done in an iterative way, rejecting points
(denoted by open circles in Fig.~\ref{fig:syniue}) located more than
$3\,\sigma$ away from the linear fit, where 
\begin{equation}
\sigma= \sqrt{{\sum (I_{syn}-b\,I_{IUE})^2}\over{(N-1)}},
\end{equation} 
and $N$ is the number of stars.

Columns 2--5 of Table~\ref{tab:calib} give the results of the calibration, its
associated error, the rms in magnitudes, and the number of stars left after
removing objects as described above.


\section{Summary}

In this work we have presented a set of seventeen mid-UV indices computed from
the synthetic stellar SEDs of the UVBLUE library. We have explored for the
first time their sensitivity in terms of the atmospheric parameters
($T_\mathrm{eff}$, $\log{g}$, [M/H]) as well as spectral resolution. We find
that theoretical indices display a wide variety of behaviors, however,
qualitatively resembling the results from empirical analysis using IUE
data. This is, all line indices peak at effective temperatures of the order of
6500~K or lower and are higher with increasing gravity. For the purpose of
analyzing integrated spectra of evolved populations the detection of features
involved in any of the indices manifests the presence of stars older that
1~Gyr.

Bearing in mind that the UV light of stellar aggregates is dominated by stars
at the turn off, the effects of chemical composition are more important than
those due to surface gravity. The results indicate that in general high
metallicity theoretical indices are larger than their metal-poor counterparts;
nevertheless for some of them (Fe~\textsc{ii}~2332, Fe~\textsc{ii}~2402 and, to
some extent, also Mg~\textsc{ii}~2800) we find that the loci of the maxima
strongly depend on metallicity. The result of this shift is that for a given
lower temperature than the one at the peak, metal poor indices are
higher. Interestingly we find that the index BL~3096 is even less sentitive to
metallicity than Fe~\textsc{i}~3000, which was empirically found to marginally
respond to changes of this parameter.

The effects of instrumental resolution vary from negligible to very
important, even if we only consider the range in resolution limited by IUE
and Kurucz low resolution fluxes (6 and 10~\AA, respectively). The index
BL~3096 can change up to 20\% in this interval.

Among the most important results of this work is that, after quantitatively
comparing theoretical and empirical indices, we have found that five indices
(although S2850 and S2850L measure the same property on the SED) agree very
well with observations on the entire range of the parameter space delimited by
the observed high-gravity stars. The indices cover both line and continuum
measurements that show distinctive characteristics in stars. This property
ensures their applicability in low as well as intermediate resolution analyses
of the mid-UV SED of evolved stellar populations.


\acknowledgements M.C. and E.B. are pleased to thank financial support from
Mexican CONACyT, via grants 36547-E and SEP-2004-C01-47904. A.B. acknowledges partial
financial support by the Italian MIUR under grant INAF PRIN/05 1.06.08.03. M.F., M.L.M. and C.M. would like to acknowledge funding from italian grants MIUR COFIN-2003028039 and PRIN-INAF 2005 (P.I. M. Bellazzini). Thanks are due to Alessandro Bressan for providing us with the definition of the 2110/2570, S2850, S2850L indices in Table~\ref{tab:indices}.



\clearpage

\clearpage

\clearpage

\clearpage

\clearpage

\clearpage

\clearpage
\begin{table}  
{\small
\begin{center}
\caption{Definition of the bands for line and continuum indices.}
\label{tab:indices}
\begin{tabular}{lcccl}
\tableline \tableline
\noalign{\smallskip}
            & Blue bandpass & Index bandpass & Red bandpass & Main species   \\
Index name  & (\AA)         & (\AA)          & (\AA)        &                \\
\noalign{\smallskip}
\tableline
\noalign{\smallskip}
Fe~\textsc{ii}~2332  & 2285 -- 2325 & 2333 -- 2359 & 2432 -- 2458 &   Fe~\textsc{ii}, Fe~\textsc{i}, Co~\textsc{i}, 
Ni~\textsc{i}   \\
Fe~\textsc{ii}~2402  & 2285 -- 2325 & 2382 -- 2422 & 2432 -- 2458 &   Fe~\textsc{ii}, Fe~\textsc{i}, Co~\textsc{i}   \\
BL 2538      & 2432 -- 2458 & 2520 -- 2556 & 2562 -- 2588 &   Fe~\textsc{i}, Fe~\textsc{ii}, Mg~\textsc{i}, 
Cr~\textsc{i}, Ni~\textsc{i}   \\
Fe~\textsc{ii}~2609  & 2562 -- 2588 & 2596 -- 2622 & 2647 -- 2673 & Fe~\textsc{ii}, Fe~\textsc{i}, Mn~\textsc{ii}   \\
BL 2720  & 2647 -- 2673 & 2713 -- 2733 & 2762 -- 2782 & Fe~\textsc{i}, Fe~\textsc{ii}, Cr~\textsc{i}   \\
BL 2740      & 2647 -- 2673 & 2736 -- 2762 & 2762 -- 2782 &  Fe~\textsc{i}, Fe~\textsc{ii}, Cr~\textsc{i}, 
Cr~\textsc{ii}   \\
Mg~\textsc{ii}~2800  & 2762 -- 2782 & 2784 -- 2814 & 2818 -- 2838 &  Mg~\textsc{ii}, Fe~\textsc{i}, Mn~\textsc{i}  \\
Mg~\textsc{i}~2852   & 2818 -- 2838 & 2839 -- 2865 & 2906 -- 2936 &  Mg~\textsc{i}, Fe~\textsc{i}, Cr~\textsc{ii}, 
Fe~\textsc{ii}   \\
Mg Wide      & 2470 -- 2670 & 2670 -- 2870 & 2930 -- 3130 &  Mg~\textsc{i}, Mg~\textsc{ii}, Fe~\textsc{i}, 
Fe~\textsc{ii},  Cr~\textsc{i}, Cr~\textsc{ii}   \\  
Fe~\textsc{i}~3000   & 2906 -- 2936 & 2965 -- 3025 & 3031 -- 3051 &  Fe~\textsc{i}, Cr~\textsc{i}, Fe~\textsc{ii}, 
Ni~\textsc{i}   \\
BL 3096      & 3031 -- 3051 & 3086 -- 3106 & 3115 -- 3155 &  Fe~\textsc{i}, Ni~\textsc{i}, Mg~\textsc{i}, 
Al~\textsc{i}   \\
2110/2570    & 2010 -- 2210 & ...  -- ...  & 2470 -- 2670 &   \\
2600--3000   & 2470 -- 2670 & ...  -- ...  & 2930 -- 3130 &   \\
2609/2660    & 2596 -- 2623 & ...  -- ...  & 2647 -- 2673 &   \\
2828/2921    & 2818 -- 2838 & ...  -- ...  & 2906 -- 2936 &   \\
S2850        & 2599 -- 2601 & ...  -- ...  & 3099 -- 3101 &   \\
S2850 L      & 2590 -- 2610 & ...  -- ...  & 3090 -- 3110 &   \\
\noalign{\smallskip}
\tableline
\end{tabular}  
\end{center} }
\end{table}  

\clearpage

\begin{deluxetable}{cccccccccccccc}
\tabletypesize{\tiny}
\rotate
\tablecolumns{14}
\tablewidth{0pc}
\tablecaption{Synthetic UVBLUE indices after degrading the SEDs at 6~\AA\ FWHM.
\label{tab:uvblueindices}}
\tablehead{
\colhead{$T_\mathrm{eff}$} & \colhead{$\log{g}$} & \colhead{[M/H]} &
\colhead{Fe~\textsc{ii}~2332} & \colhead{Fe~\textsc{ii}~2402} & \colhead{BL 2538} & \colhead{Fe~\textsc{ii}~2609} & \colhead{BL 2720} & 
\colhead{BL 2740} & \colhead{Mg~\textsc{ii}~2800} & \colhead{Mg~\textsc{i}~2852} & \colhead{Mg Wide} & \colhead{Fe~\textsc{i}~3000} & 
\colhead{BL 3096}  \\
\colhead{(K)} & \colhead{(dex)} & \colhead{(dex)} &
(mag) & (mag) & (mag) & (mag) & (mag) & (mag) & (mag) & (mag) & (mag) & (mag) & (mag)  }
\startdata
 4000 & 1.0 & -2.0 &  0.44061 &  0.47539 &  0.43790 &  0.87782 &  0.59571 &  0.50926 &  0.76453 &  1.35895 &  
1.04094 &  0.35647 &  0.16311   \\
 4000 & 1.5 & -2.0 &  0.41748 &  0.48331 &  0.59642 &  1.04292 &  0.67021 &  0.58925 &  0.83816 &  1.53405 &  
0.99007 &  0.39780 &  0.20965   \\
 4000 & 2.0 & -2.0 &  0.37282 &  0.53914 &  0.77864 &  1.11860 &  0.76399 &  0.67206 &  0.88320 &  1.70192 &  
0.89574 &  0.44023 &  0.29045   \\
 4000 & 2.5 & -2.0 &  0.29809 &  0.61892 &  0.93896 &  1.08568 &  0.85078 &  0.73388 &  0.85145 &  1.75295 &  
0.76781 &  0.47426 &  0.40146   \\
 4000 & 3.0 & -2.0 &  0.23567 &  0.62265 &  0.98577 &  0.91336 &  0.83173 &  0.69016 &  0.70402 &  1.65113 &  
0.62329 &  0.45628 &  0.46000   \\
 4000 & 3.5 & -2.0 &  0.17223 &  0.55195 &  0.98687 &  0.68457 &  0.77044 &  0.60286 &  0.52936 &  1.50798 &  
0.50655 &  0.40967 &  0.46412   \\
 4000 & 4.0 & -2.0 &  0.09713 &  0.48134 &  0.99168 &  0.50300 &  0.72933 &  0.54143 &  0.38437 &  1.35511 &  
0.43448 &  0.37278 &  0.45930   \\
 4000 & 4.5 & -2.0 & -0.21163 &  0.50405 &  0.88451 &  0.80879 &  1.23938 &  0.77861 &  0.11611 &  0.86800 &  
0.48530 &  0.26252 &  0.39315   \\
 4000 & 5.0 & -2.0 & -0.50491 &  0.36852 &  0.94043 &  0.43989 &  1.21883 &  0.77068 &  0.07321 &  0.85313 &  
0.53453 &  0.28996 &  0.47198   \\
 4500 & 1.5 & -2.0 &  0.31132 &  0.48188 &  0.18393 &  0.41794 &  0.11088 &  0.21140 &  0.59527 &  0.13673 &  
0.26370 &  0.05283 &  0.08340   \\
 4500 & 2.0 & -2.0 &  0.71607 &  1.10758 &  0.53916 &  1.37005 &  0.34172 &  0.51743 &  1.31980 &  0.89645 &  
0.48798 &  0.18059 &  0.18137   \\
 4500 & 2.5 & -2.0 &  0.61902 &  1.07981 &  0.67985 &  1.28936 &  0.41383 &  0.53511 &  1.30672 &  1.17151 &  
0.43161 &  0.22482 &  0.21524   \\
 4500 & 3.0 & -2.0 &  0.51926 &  1.02658 &  0.83930 &  1.22573 &  0.49990 &  0.56285 &  1.25480 &  1.46864 &  
0.39972 &  0.28141 &  0.26612   \\
 4500 & 3.5 & -2.0 &  0.42160 &  0.96628 &  1.01393 &  1.17014 &  0.59235 &  0.59360 &  1.15795 &  1.69445 &  
0.38010 &  0.34159 &  0.33579   \\
\enddata
\tabletypesize{\small}
\tablecomments{Table \ref{tab:uvblueindices} is published in its entirety in the 
  electronic edition of the {\it Astrophysical Journal}.  A portion is 
  shown here for guidance regarding its form and content.}
\end{deluxetable}

\clearpage

\begin{table}
\caption{Transformation coefficients.}
\label{tab:calib}
\begin{tabular}{lcccc}

\tableline \tableline
\noalign{\smallskip}
Index name  & b         & $\sigma_b$ & rms   & \#  \\
            &           &            & (mag) &     \\
\noalign{\smallskip}
\tableline
\noalign{\smallskip}
BL 2538      &  1.517 &  0.041 &  0.18 &  62 \\
BL 2720      &  3.028 &  0.117 &  0.12 &  67 \\
Mg~\textsc{i}~2852   &  1.797 &  0.036 &  0.13 &  66 \\
Mg Wide      &  0.811 &  0.013 &  0.03 &  64 \\
Fe~\textsc{i}~3000   &  0.966 &  0.025 &  0.04 &  66 \\
BL 3096      &  1.260 &  0.046 &  0.05 &  66 \\
2110/2570    &  0.934 &  0.027 &  0.24 &  60 \\
2600--3000   &  0.717 &  0.012 &  0.14 &  67 \\
2828/2921    &  1.068 &  0.021 &  0.08 &  60 \\
S2850        &  1.014 &  0.011 &  0.14 &  64 \\
S2850 L      &  0.924 &  0.010 &  0.13 &  67 \\
\noalign{\smallskip}
\tableline
\end{tabular}
\end{table}


\end{document}